\documentclass[%
notitlepage,
superscriptaddress,pre,showpacs,twocolumn,aps]{revtex4-1}
\usepackage{natbib}
\usepackage{subfigure}
\usepackage{graphicx}
\usepackage{dcolumn}
\usepackage{bm}
\usepackage{epstopdf}
\usepackage{epsfig}
\usepackage{amsmath,amssymb}
\usepackage[colorlinks = true,
linkcolor = blue,
urlcolor  = blue,
citecolor = blue,
anchorcolor = blue]{hyperref}
\usepackage{url}

\DeclareMathOperator{\csch}{csch}

\newcommand{\pts}{$\cal{PT}$}

\begin{document}
	
	
	\title{Snakes and ghosts in a parity-time-symmetric chain of dimers}
	
	\author{H. Susanto}
	\email{hsusanto@essex.ac.uk}
	\affiliation{%
		Department of Mathematical Sciences, University of Essex, Wivenhoe Park, Colchester CO4 3SQ, United Kingdom
	}%

	\author{R. Kusdiantara}%
	\affiliation{%
		Department of Mathematical Sciences, University of Essex, Wivenhoe Park, Colchester CO4 3SQ, United Kingdom
	}%
	\affiliation{Centre of Mathematical Modelling and Simulation, Institut Teknologi Bandung, 1st Floor, Labtek III, Jl.\ Ganesha No.\ 10, Bandung, 40132, Indonesia}

	\author{N.\ Li}
	\affiliation{%
		School of Computer Science and Electronic Engineering, University of Essex, Wivenhoe Park, Colchester CO4 3SQ, United Kingdom
	}%
	
	\author{O.B. Kirikchi}
	\affiliation{%
		Department of Mathematical Sciences, University of Essex, Wivenhoe Park, Colchester CO4 3SQ, United Kingdom
	}%
	
	\author{D.\ Adzkiya}
	\affiliation{Department of Mathematics, Faculty of Mathematics and Natural Sciences,
		Institut Teknologi Sepuluh Nopember, Sukolilo, Surabaya 60111, Indonesia}
	
	\author{E.R.M. Putri}
	\affiliation{Department of Mathematics, Faculty of Mathematics and Natural Sciences,
	Institut Teknologi Sepuluh Nopember, Sukolilo, Surabaya 60111, Indonesia}

\author{T. Asfihani}
	\affiliation{Department of Mathematics, Faculty of Mathematics and Natural Sciences,
	Institut Teknologi Sepuluh Nopember, Sukolilo, Surabaya 60111, Indonesia}

	\date{\today}
	
	\begin{abstract}
		We consider linearly coupled discrete nonlinear Schr\"odinger equations with gain and loss terms and with a cubic-quintic nonlinearity. The system models a parity-time ($\cal{PT}$)-symmetric coupler composed by a chain of dimers. Particularly we study site-centered and bond-centered spatially-localized solutions and present that each solution has a symmetric and antisymmetric configuration between the arms. When a parameter is varied, the resulting bifurcation diagrams for the existence of standing localized solutions have a snaking behaviour.  The critical gain/loss coefficient above which the $\cal{PT}-$symmetry is broken corresponds to the condition when bifurcation diagrams of symmetric and antisymmetric states merge. Past the symmetry breaking, the system no longer has time-independent states. Nevertheless, equilibrium solutions can be analytically continued by defining a dual equation that leads to so-called ghost states associated with growth or decay, that are also identified and examined here. We show that ghost localized states also exhibit snaking bifurcation diagrams. We analyse the width of the snaking region and provide asymptotic approximations in the limit of strong and weak coupling where good agreement is obtained. 
	\end{abstract}
	
	\maketitle
	
	
\section{Introduction}

Many nonlinear dynamical systems, such as spatially extended nonlinear dissipative systems \cite{Purwins2010}, vertical-cavity semiconductor optical amplifiers \cite{Barbay2008}, nematic liquid crystal layers with spatially modulated input beam \cite{Haudin2011}, and magnetic fluids \cite{Lloyd2015}, exhibit spatially localized patterns and a snaking structure in their  bifurcation diagrams in the plane of the length of the localized solution against a control parameter. This phenomenon of snaking is referred to as homoclinic snaking \cite{Pomeau1986,coul00,Woods2006}, where the spatial structure of such a localized state departs from and then returns to a uniform state. By definition, it has infinitely many turning points (i.e.\ saddle-node or saddle-centre bifurcations), which form the boundaries of the snaking region. Such a region is also called pinning region since the fronts at either end `pin' or `lock' to the structure within the localized state. An infinite number of localized states exist 
in the entire interval of the pinning region. 

In most of previous works devoted to localized states and snaking in continuous systems, the Swift-Hohenberg equation has been widely used as a model for pattern formation since it is the simplest model equation that illustrates the pinning effect \cite{Sakaguchi1996,Woods2006,Burke2006,Burke2007,Burke2007a}. In general, the effect cannot be described by conventional multiple-scale asymptotic method due to the fact that the length of the pinning region is exponentially small in a parameter which is related to the pattern amplitude \cite{Pomeau1986}. Recently the Swift-Hohenberg with quadratic-cubic nonlinearities and cubic-quintic nonlinearities have been successfully studied with the help of exponential asymptotics \cite{Chapman2009,Kozyreff2006,Dean2011}. The calculations, however, are rather cumbersome and require two fitting parameters. Alternatively, variational methods to obtain scaling laws for the structure of the snaking region have been proposed and demonstrated, for example, in the system modelled by the cubic-quintic Swift-Hohenberg equation \cite{Susanto2011}.

Like spatially continuous systems, several discrete systems can display the snaking behavior with the locking effect, however, being attributed to the imposed lattice. Examples include the discrete bistable nonlinear Schr{\"o}dinger equation \cite{Carretero-Gonzalez2006,Chong2009,Chong2011}, which leads to a subcritical Allen-Cahn equation \cite{Taylor2010}, optical cavity solitons \cite{Yulin2008,Yulin2010}, discrete systems with a weakly broken pitchfork bifurcation \cite{Clerc2011} and in patterns on networks appearing due to Turing instabilities \cite{mccu16}. Pinning regions in lattices were studied analytically by Matthews and Susanto \cite{Matthews2011} and Dean et al.\ \cite{Dean2015}.

This paper is devoted to a detailed numerical and analytical study of homoclinic snaking in a parity-time ($\mathcal{PT}$) system. The physical problem is a chain of dimers that has two arms with each arm described by a discrete nonlinear Schr{\"o}dinger equation with gain or loss and with cubic-quintic nonlinearity. While the concept of $\mathcal{PT}-$symmetry has gained a lot of attention in the last decade \cite{such16,kono16}, to the best of our knowledge, the effect of the gain and loss term in a $\mathcal{PT}$-symmetric chain of dimers to the snaking regime has not been explored yet.

A system of equations is $\mathcal{PT}-$symmetric when it is invariant with respect to combined parity ($\mathcal{P}$) and time-reversal ($\mathcal{T}$) transformation \cite{bend98,bend99,bend07}. In the context of Schr\"odinger Hamiltonians with a complex potential $V(x)$, $\mathcal{PT}-$symmetry requires the potential to satisfy the condition $V(x)=V^*(-x)$, where $^*$ is the complex conjugation, i.e., $V(x)$ has a symmetric real part and an antisymmetric imaginary part. Such symmetry is of great interest as it forms a particular class of widely studied non-Hermitian Hamiltonians in quantum mechanics \cite{mois11}, that does not satisfy the standard postulate that the Hamiltonian operator be Dirac Hermitian and yet can have real eigenvalues up to a critical value of the complex potential parameter. Above the value, the symmetry is broken, i.e.\ the eigenvalues of the Hamiltonian become complex-valued. Among $\mathcal{PT}-$symmetric systems, dimers are the most basic and important. The concept was first demonstrated experimentally on dimers, which are composed of two coupled optical waveguides \cite{guo09,rute10} (see also \cite{pick13} and references therein). In particular, when nonlinear dimers are put in arrays where elements with gain and loss are linearly coupled to the elements of the same type belonging to adjacent dimers, one can obtain a distinctive feature in the form of the existence of solutions localized in space as continuous families of their energy parameter \cite{such11}. The nonlinear localized solutions and their stability have been studied in \cite{kiri16,alex12,alex17} analytically and numerically (see also the references therein for localized solutions in systems of coupled nonlinear Schr\"odinger equations).

The continuum limit of the set-up studied herein was considered in \cite{burl16,burl13}. In optical media, such nonlinearity can be obtained from a saturation of the Kerr response, which with the increase of the intensity will introduce a self-defocusing quintic term in the expansion of the refractive index \cite{cout91,smir06}. In the continuous case \cite{burl16,burl13}, it was shown that the presence of gain/loss terms only influences the stability of the localized solutions. Here, it will be shown that the discrete set-up admits homoclinic snaking. We show that the critical gain/loss parameter corresponding to the `broken \pts\ symmetry' phase is related to the merging of two snaking regions. Beyond the critical point, the system does not have time-independent states. Nevertheless, their continuation can be analytically obtained by defining a dual system. Here, we also identify and examine localized solutions of the dual equations, where interestingly we obtain that they also preserve the snaking region, including its width.

The report is outlined as follows. The $\mathcal{PT}$-symmetric chain of dimers with cubic-quintic nonlinearity is discussed in Section \ref{sec2}. In Section \ref{sec2b}, we study spatially uniform solutions and their stability. We obtain that symmetric states can become unstable due to pitchfork bifurcations. The emanating solutions are asymmetric. In the presence of gain/loss parameter, such solutions are lost. By setting a complex-valued propagation parameter, they can be recovered. However, they are not actual solutions of the governing equations and are referred to as ghost states, that are discussed in Section \ref{sec2c}. We study localized solutions, their stability, and the observation of homoclinic snaking numerically in Section \ref{sec3}. When the \pts-symmetry is broken, we can also define ghost states as continuations of uniform and localized solutions discussed in the previous sections. We analyse these states in Section \ref{sec3b}. Section \ref{sec4} is on the asymptotic expression of the snaking width that is obtained in the limit of small and large coupling, which is then compared with computational results, where good agreement is obtained. In the strong coupling region, we use a variational method following \cite{Matthews2011}, but with a different approach yielding a simple expression of the width that was not obtainable in \cite{Matthews2011}. In the weak coupling case, we introduce a one-active-site approximation following \cite{kusd16}. Conclusions are given in Section \ref{sec5}.

\section{Mathematical model and stability of solutions}
\label{sec2}

The governing equations describing \pts-symmetric chains of dimers are of the form
\begin{equation}
\begin{aligned}
i\dot{u}_n&=\left(C\Delta_{2}-\omega + \left|u_n\right|^2 -Q\left|u_n\right|^4 +i\gamma\right) u_n+v_n,\\
i\dot{v}_n&=\left(C\Delta_{2}-\omega + \left|v_n\right|^2 -Q\left|v_n\right|^4  -i\gamma\right) v_n+u_n.
\end{aligned}
\label{gov1}
\end{equation}
The derivative with respect to the evolution variable (i.e., the propagation distance, if we consider their application in fiber optics) is denoted by the overdot, $u_{n}=u_{n}(t)$, $v_{n}=v_{n}(t)$  are complex-valued wave function at site $n\in\mathbb{Z}$ with the propagation constant $\omega\in\mathbb{R}$, $C>0$ is the constant coefficient of the horizontal linear coupling (coupling constant between two adjacent sites), $\Delta_{2}\square_{n}=(\square_{n+1}-2\square_{n}+\square_{n-1})$ 
is the discrete Laplacian term in one spatial dimension, and the gain and loss acting on complex variables $u_{n}$, $v_{n}$ are represented by the parameter $\gamma> 0$. The cubic nonlinearity coefficient has been scaled to $+1$, while $Q$ is the coefficient of the quintic nonlinearity. 

System \eqref{gov1} is $\mathcal{PT}$-symmetric because it is invariant with respect to the action of the parity $\mathcal{P}$ and time-reversal $\mathcal{T}$ operators
given by 
\begin{align}
\mathcal{P}\begin{pmatrix}
u_n(t)\\v_n(t)
\end{pmatrix}
=\begin{pmatrix}
v_n(t)\\u_n(t)
\end{pmatrix},\, 
\mathcal{T}\begin{pmatrix}
u_n(t)\\v_n(t)
\end{pmatrix}
=\begin{pmatrix}
u_n^*(-t)\\v_n^*(-t)
\end{pmatrix}.
\end{align}

Next, we consider the equations for standing wave solutions of Eqs.\ \eqref{gov1}, obtained from setting $\dot{u}_n=\dot{v}_n=0$ and substituting $u_{n}=A_{n},\,v_{n}=B_{n}e^{i\phi}$ into (\ref{gov1}), 
\begin{equation}
\begin{aligned}
&\left(C\Delta_2 -\omega+ A_n^2-Q A_n^4+i\gamma\right) A_n+B_n e^{i\phi}=0,\\
&\left(C\Delta_{2}-\omega+ B_n^2-Q B_n^4-i\gamma\right) B_n+A_n e^{-i\phi}=0.
\end{aligned}
\label{gov2}
\end{equation}
Here, $A_n,B_n,\phi\in\mathbb{R}$. We can assume that $u_n$ is real-valued because of the phase invariance of the governing equations \eqref{gov1}. Splitting the real and imaginary parts of the equations and simplifying them will yield
\begin{equation}
\begin{aligned}
\Omega A_n =& C\Delta_2A_n+ A_n^3-Q A_n^5,
\end{aligned}
\label{gov3}
\end{equation}
which is also known as the discrete Allen-Cahn equation, where $B_n=A_n$, $\Omega=\omega\mp\sqrt{1-\gamma^2}$ with $\phi=-\arcsin\gamma$ for the minus sign and $\phi=\pi+\arcsin\gamma$ for the plus sign, which corresponds to the so-called symmetric and antisymmetric configuration between the arms, respectively. Note that (\ref{gov3}) will have no real solution when $\gamma>1$. This is the broken region of \pts-symmetry. 

The linear stability of a standing wave solution is determined as follows. 
Introducing the ansatz $u_{n}=A_{n}+{\epsilon}(\widetilde{u}_{r,n}+i \widetilde{u}_{i,n})e^{\lambda t}$, $v_{n}=B_{n}e^{i\phi}+{\epsilon}(\widetilde{v}_{r,n}+i \widetilde{v}_{i,n})e^{\lambda t}$, $|{\epsilon}| \ll 1$, and substituting it into Eq.\ (\ref{gov1}) will yield from the terms at $\mathcal{O}({\epsilon}$) the linear eigenvalue problem 
\begin{equation}
\begin{aligned}
\lambda{\widetilde{u}_{r,n}}&=(C\Delta_2 - \omega+A_n^2-QA_n^4)\widetilde{u}_{i,n}+\gamma \widetilde{u}_{r,n}+\widetilde{v}_{i,n},\\
-\lambda{\widetilde{u}_{i,n}}&=(C\Delta_2-\omega+3A_{n}^{2}-5QA_n^4)\widetilde{u}_{r,n}-\gamma \widetilde{u}_{i,n}+\widetilde{v}_{r,n},\\
\lambda{\widetilde{v}_{r,n}}&=(C\Delta_2 - \omega +\tilde{B}_n^2-Q\tilde{B}_n^4)\widetilde{v}_{i,n}\\
&+(2\gamma B_n\tilde{B}_n(1-2Q\tilde{B}_n^2)-\gamma)\widetilde{v}_{r,n}+\widetilde{u}_{i,n},\\
-\lambda \widetilde{v}_{i,n}&=(C\Delta_2 -\omega+ 3\tilde{B}_n^2-5Q\tilde{B}_n^4)\widetilde{v}_{r,n}+\gamma \widetilde{v}_{i,n}+\widetilde{u}_{r,n},
\end{aligned}
\label{evp}
\end{equation}
where $\tilde{B}_n=B_n\sqrt{1-\gamma^2}$. 

Generally the spectrum will consist of two types, i.e.\ continuous and discrete spectrum or eigenvalue. A solution is unstable when there exists $\lambda$ with Re$(\lambda)> 0$. However, if $\lambda$ is a spectrum, so are $-\lambda$ and $\pm\overline{\lambda}$ \cite{kiri16}. A solution is therefore (linearly) stable only when Re$(\lambda)= 0$ for all $\lambda$, i.e.\ it is neutral stability. Nonlinear stability may be obtained numerically by evolving a perturbed solution in Eqs.\ \eqref{gov1} for a long while, which analytically 
is still an open problem due to the absence of a Hamiltonian structure of the system (see, e.g., \cite{cher16,dest17} for nonlinear stability analysis of a similar system but with cross-dispersion and different nonlinearity that becomes possible because it has a Hamiltonian form via a cross-gradient symplectic structure).

Numerically we solve the steady-state equations of \eqref{gov2} 
using a Newton-Raphson method in \textsc{Matlab}. A pseudo-arclength continuation scheme is implemented to do numerical continuation past a turning point. To model the infinite domain, we use a periodic boundary condition with a large number of lattices. The typical value we use is $N=100$, but larger values were used as well to guarantee that the results are independent of the number of sites. After a solution is obtained, its stability is determined numerically by solving Eqs.\ \eqref{evp} using a standard eigenvalue problem solver. 

\begin{figure}[thbp!]
	\centering
	\subfigure[]{\includegraphics[scale=0.52]{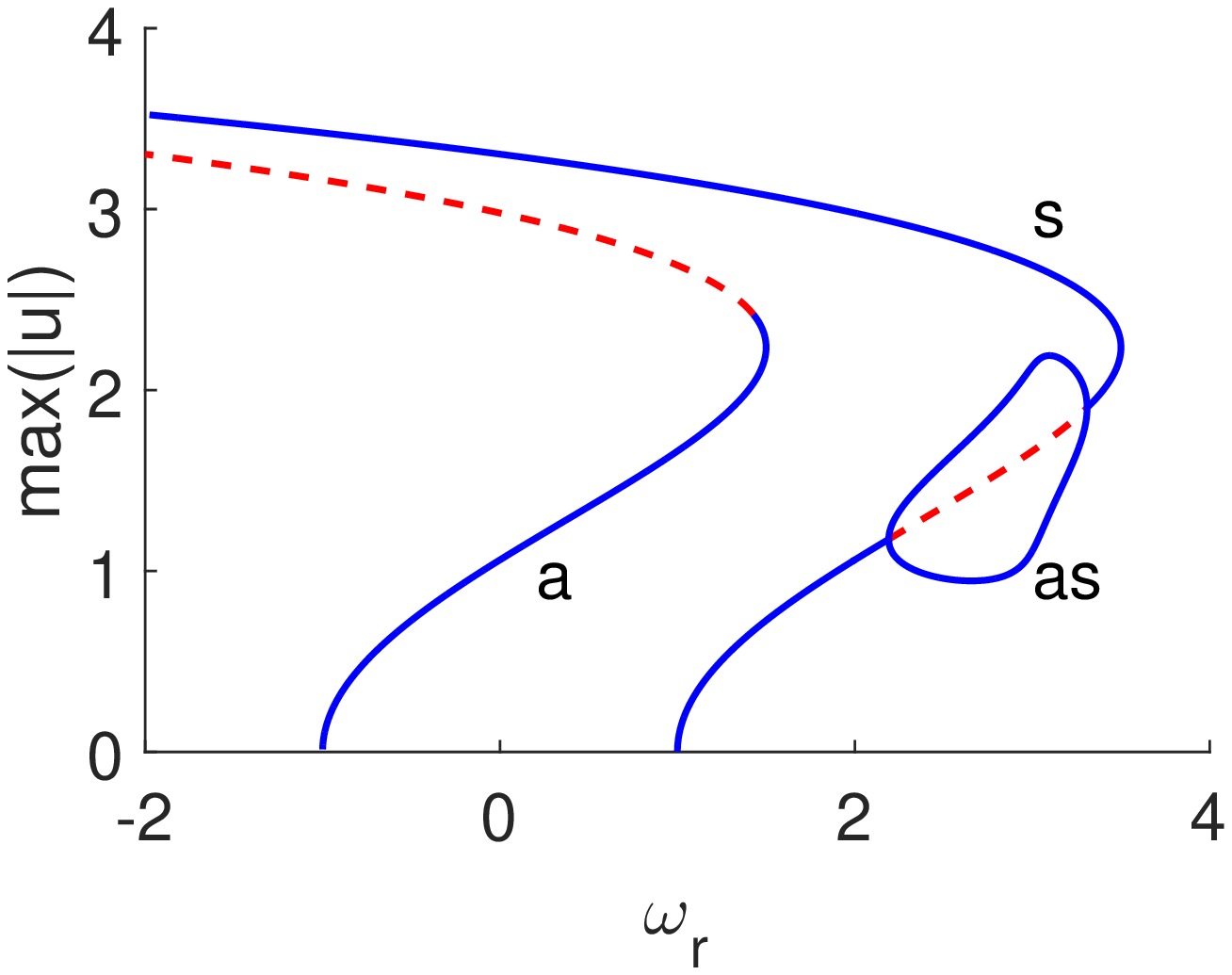}}
	\subfigure[]{\includegraphics[scale=0.52]{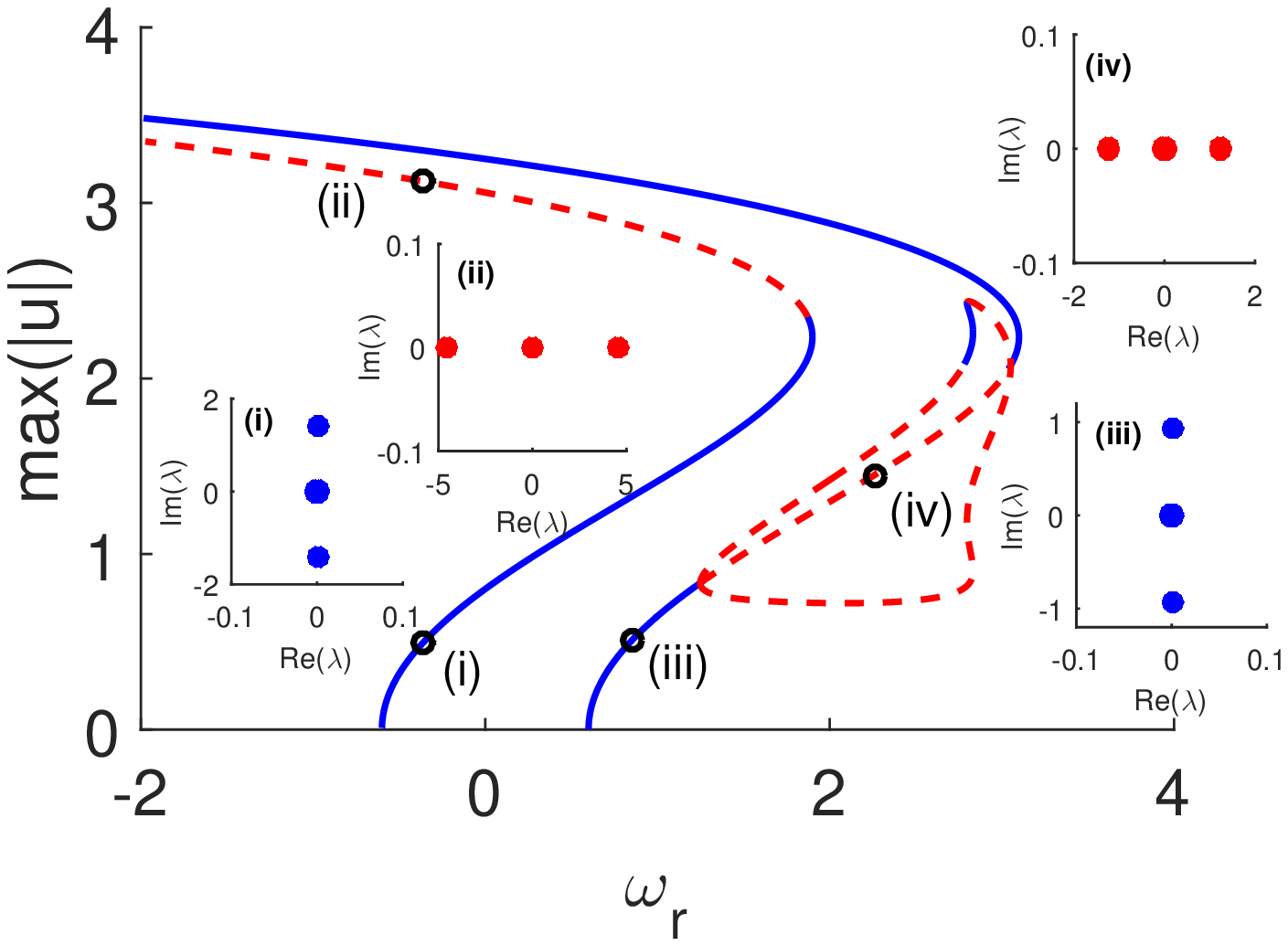}}
	\subfigure[]{\includegraphics[scale=0.52]{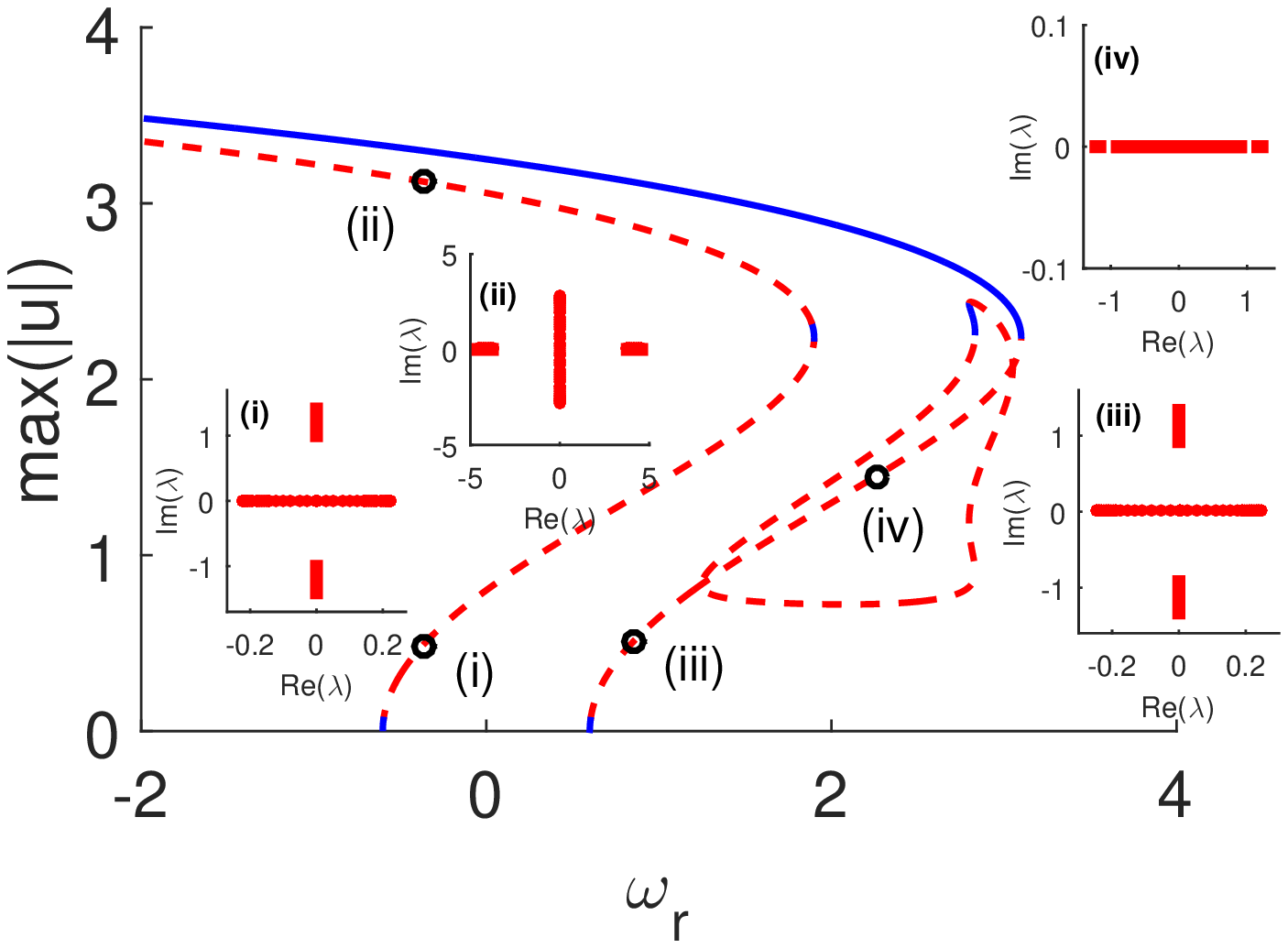}}
	\caption{Bifurcation diagram of the equilibria of \eqref{gov1} for (a) {$C=\gamma=0$, (b) $C=0$ and $\gamma=0.8$}, (c) {$C=0.1$ and $\gamma=0.8$}. Here, $Q=0.1$ and $\omega=\omega_r+i\omega_i$. Branches 'a' and 's' are antisymmetric 
		and symmetric 
		configuration between the arms, respectively, obtained from \eqref{usol}, i.e., $u=A$. Along the branches, $\omega_i=0$. Branch 'as' corresponds to asymmetric solutions, obtained from the nullclines of \eqref{gov1} with $\omega_i$ given by \eqref{omi}. 
Stable solutions are shown as solid line and as dashed line otherwise. The insets are linear spectrum of the indicated solutions.}
	\label{fig:uniform_ghost}		
\end{figure}

\section{Uniform solutions}
\label{sec2b}

Equation \eqref{gov3} has uniform solutions $A_n\equiv A$ that are given by
\begin{align}
A=0,\,A^2=\frac{1\pm\sqrt{1-4Q\Omega}}{2Q}.
\label{usol}
\end{align}
Besides $\gamma$ must be less than $1$, the uniform solution \eqref{usol} also requires $4Q\Omega<1$ to exist. Under competing cubic-quintic nonlinearities, i.e., $Q>0$, we will have two branches of non-zero uniform solutions. 

The stability of uniform solutions (\ref{usol}) can be determined by computing their continuous spectrum. Introducing the plane-wave ansatz
$(\widetilde{u}_{r,n},\widetilde{u}_{i,n},\widetilde{v}_{r,n}.\widetilde{v}_{i,n})=(\hat{k},\hat{l},\hat{p},\hat{q})e^{ik{n}}
$, $k\in\mathbb{R}$, and substituting it into (\ref{evp}) will yield a dispersion relation. Continuous spectrum of the equilibrium is then obtained by setting $k=0$ and $k=\pi$ in the dispersion equation. 

The dispersion relation of the trivial equilibrium $A_n=B_n=0$ is 
\begin{eqnarray}
\lambda^2=&
-\left(K/2-\omega\mp\sqrt{1-\gamma^2}\right)^2,
\label{dps}
\end{eqnarray}
with $K=4C(\cos k -1)$, from which we obtain the continuous spectrum 
$\lambda \in\pm i[\lambda_{1{-}}, \lambda_{2{-}}]$ and $\lambda \in\pm i[\lambda_{1{+}}, \lambda_{2{+}}]$  with the spectrum boundaries
\begin{eqnarray}
\lambda_{1{\pm}}&=&
\sqrt{1-\gamma^2}\pm\omega,
\label{l1}\\
\lambda_{2{\pm}}&=&\sqrt{1-\gamma^2+(\omega+4C)^2-2\sqrt{1-\gamma^2}(4C\mp\omega)}.\label{l2}
\end{eqnarray}
The equilibrium is therefore stable for $0<\gamma^2<1$ and unstable otherwise. Continuous spectra of the non-zero solution can be obtained similarly. 

When $C=0$, bifurcation diagrams of the nonzero solutions 
are shown in Fig.\ \ref{fig:uniform_ghost}(a,b) for two values of $\gamma$. In this case, the chain is uncoupled and one obtains the dimer, which was studied in \cite{pick13} (see also references therein) for $Q=0$ and in \cite{li17} for nonzero $Q$. 

Consider antisymmetric solutions along branch 'a'. We obtain that the low intensity solution, i.e.\ the lower branch, is stable, while the high one is not. Branch 's' generally corresponds to stable symmetric solutions, but there is a small portion of unstable branch due to pitchfork (i.e.\ spontaneous symmetry breaking) bifurcations. Solutions emanating from the branching points are asymmetric, i.e.\ $|u_n|\neq|v_n|$, and denoted by branch 'as' in Fig.\ \ref{fig:uniform_ghost}. However, they cannot be obtained from Eq.\ \eqref{usol} because they do not satisfy the parity ($\mathcal{P}$) symmetry. These will be discussed in the Section \ref{sec2c} below.

In panel (b), we consider $\gamma=0.8$. As the gain/loss parameter increases towards the critical value $\gamma=1$, branches 's' and 'a' 
become closer to each other. At the critical value, the two branches coincide, i.e.\ we obtain a turning point. This is due to the fact that when studying time-independent solutions, the governing equation \eqref{gov1} reduces nicely to the discrete Allen-Cahn equation \eqref{gov3} that is rather independent of $\gamma$ and at the same time, $\Omega$ for the symmetric and antisymmetric solutions becomes equal at $\gamma=1$.

Panel (c) shows the effect of coupling constant that clearly only affects the stability of the equilibrium. Now we obtain that branch 'a' and the lower part of branch 's' have become unstable.

\section{Asymmetric solutions as ghost states}
\label{sec2c}

In the classical cubic dimer, i.e., Eqs.\ \eqref{gov1} with $C=Q=\gamma=0$, symmetric solutions are known to be unstable for $\omega>2$ due to a pitchfork (symmetry-breaking) bifurcation (see, e.g., \cite{smer97,ragh99,kirr08,kirr11,jian14,rodr13} and references therein). At the bifurcation point, an asymmetric state pair emanate. It is a matter of a standard perturbation expansion that for $0<Q\ll1$, asymmetric solutions will bifurcate at $\omega=2+Q+\mathcal{O}(Q^2)$. This is in agreement with the result in Fig.\ \ref{fig:uniform_ghost}(a). The bifurcation diagram of the asymmetric states denoted as branch 'as' can simply be obtained from \eqref{gov1} with $\gamma=0$. However, in the cubic-quintic dimer they only exist in a finite interval. Even for larger $Q$, they may not exist at all, i.e.\ the symmetric states can be stable in their entire existence region. 

When $\gamma\neq0$, symmetric solutions still can become unstable, but the bifurcating asymmetric ones will no longer exist \cite{pick13,rodr13}. This observation was first reported in \cite{hill06}. Cartarius et al.\ \cite{cart12} provide an analytic continuation of the asymmetric solutions that emerge as ghost states, namely, a solution of the steady-state problem with complex (instead of real) valued parameter $\omega$ and hence is not a true solution of the original system \eqref{gov1}. Rodrigues et al.\ \cite{rodr13} used the proposal to obtain continuation of asymmetric states in a nonlinear Schr\"odinger equation with $\mathcal{PT}-$symmetric double-well potentials and the two-mode reduction in the form of a cubic dimer, i.e.\ \eqref{gov1} with $Q=0$. 

To obtain asymmetric states of our problem, consider again time-independent equations of Eqs.\ \eqref{gov1} and their conjugates, where the propagation constant $\omega$ is now complex-valued, i.e., $\omega=\omega_r+i\omega_i$, 
\begin{subequations}
	\begin{align}
&\left(C\Delta_{2}-\omega + \left|u_n\right|^2 -Q\left|u_n\right|^4 +i\gamma\right) u_n+v_n=0,\label{ea}\\
&\left(C\Delta_{2}-\omega + \left|v_n\right|^2 -Q\left|v_n\right|^4  -i\gamma\right) v_n+u_n=0,\label{eb}\\
&\left(C\Delta_{2}-\omega^* + \left|u_n\right|^2 -Q\left|u_n\right|^4 -i\gamma\right) u_n^*+v_n^*=0,\label{ec}\\
&\left(C\Delta_{2}-\omega^* + \left|v_n\right|^2 -Q\left|v_n\right|^4  +i\gamma\right) v_n^*+u_n^*=0.\label{ed}
\end{align}
\end{subequations}
The imaginary part $\omega_i$ needs to be determined from a consistency equation below. 

Multiplying Eqs.\ \eqref{ea}-\eqref{ed} with $u_n^*$, $v_n^*$, $-u_n$, and $-v_n$, respectively, summing up the infinite-dimensional vectors over $n$, and adding the resulting equations will lead to the equation for $\omega_i$:
\begin{equation}
\displaystyle\omega_i=\frac{\gamma\sum_n 
	\left(|u_n|^2-|v_n|^2\right)}{\sum_n |u_n|^2+|v_n|^2}.
\label{omi}
\end{equation}
It is clear that $\omega_i$ will vanish either when $|u_n|=|v_n|$, i.e., symmetric and antisymmetric solutions, or when $\gamma=0$. 

In Fig.\ \ref{fig:uniform_ghost}(b,c) the branch of asymmetric solutions is obtained from time-independent equations of \eqref{gov1} with \eqref{omi}. We also have determined the states' stability by solving the corresponding linear eigenvalue problem even though they are not actual solutions of the original system \eqref{gov1}. 

\section{Localised solutions}
\label{sec3}

\begin{figure}[htpb]
	\centering
	\subfigure[]{\includegraphics[scale=0.55]{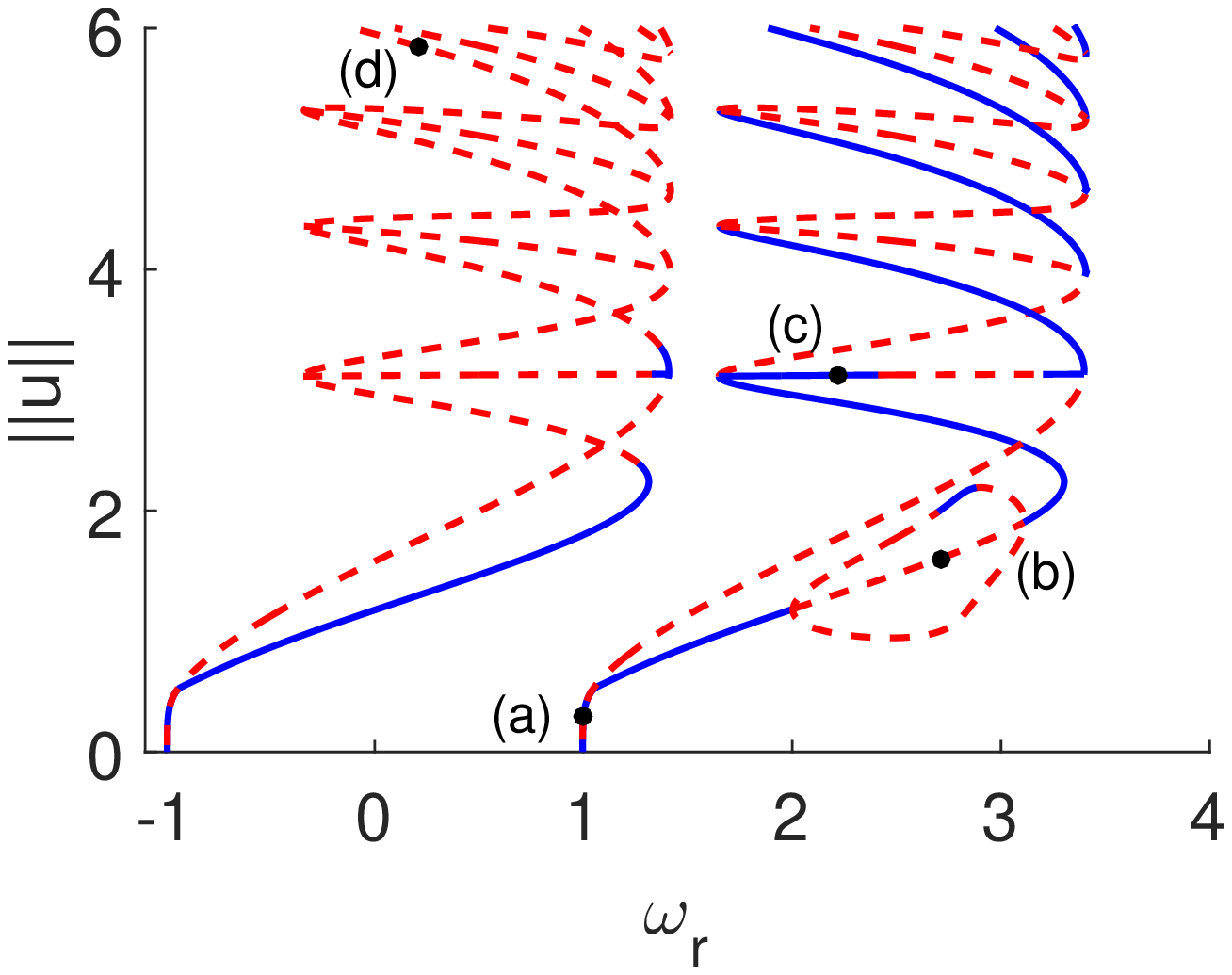}\label{subfig:bifur_c_0_1_gamma_0_1_snake_only}}
	\subfigure[]{\includegraphics[scale=0.55]{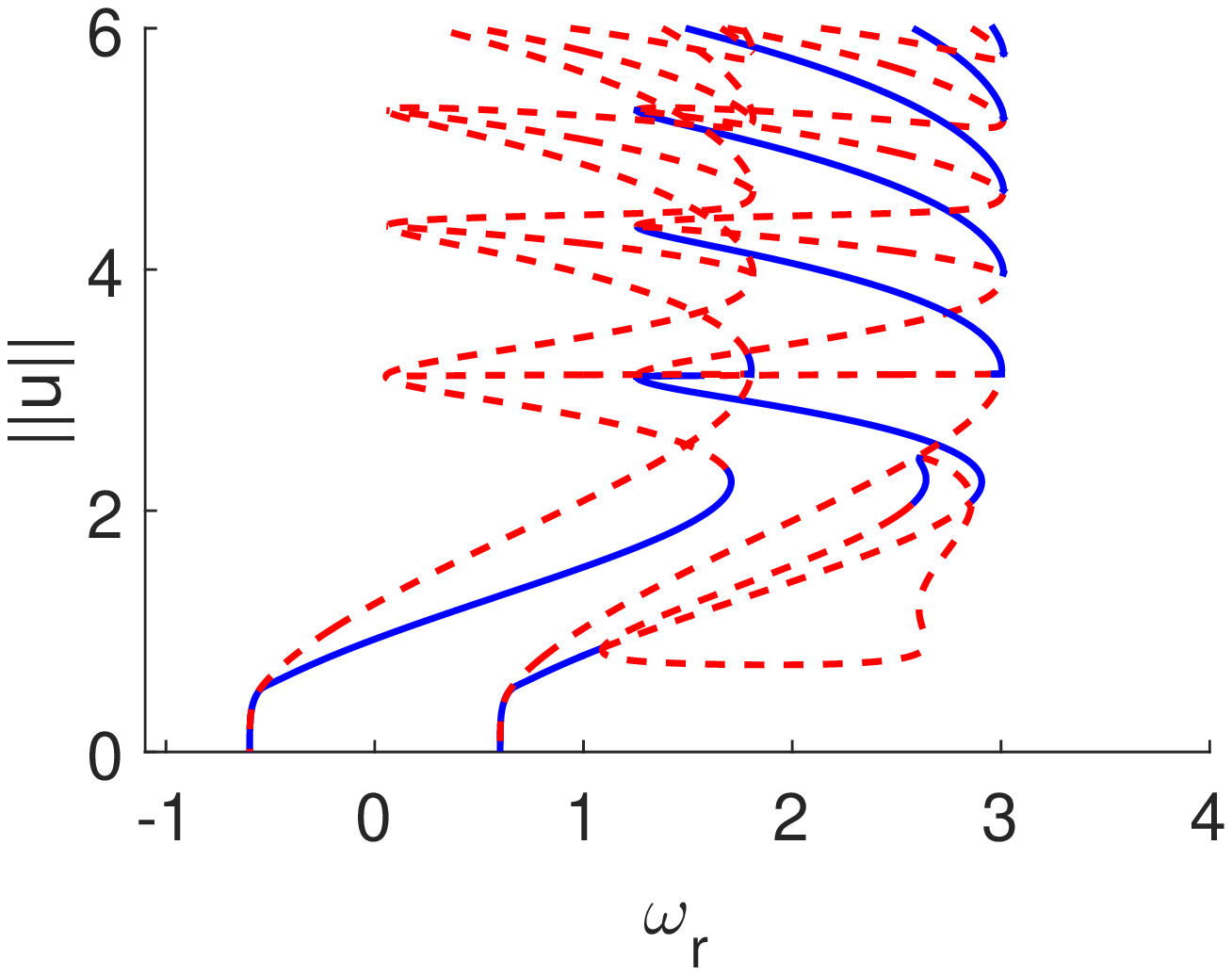}\label{subfig:bifur_c_0_1_gamma_0_8_snake_only}}
	\caption{ Bifurcation diagrams of fundamental localized solutions of Eqs.\ \eqref{gov1} {for (a) $\gamma=0.1$ and (b) $\gamma=0.8$ with $C=0.1$}. 
		Again, $\omega=\omega_r+i\omega_i$. We plot the norm $||u||=\left(\sum_n u_n^2\right)^{1/2}$ for varying $\omega_r$. There are two pairs of snaking principle branches. Each pair is connected by 'ladders' of asymmetric solutions along the same arms. Except along the closed curve of asymmetric states between the arms (that looks like branch 'as' in Fig.\ \ref{fig:uniform_ghost}), $\omega_i=0$. Solutions at indicated points in panel (a) are plotted in Fig.\ \ref{fig:prof}.}
	\label{fig2}
\end{figure}

\begin{figure*}[thbp!]
	\centering
	\subfigure[]{\includegraphics[scale=0.55]{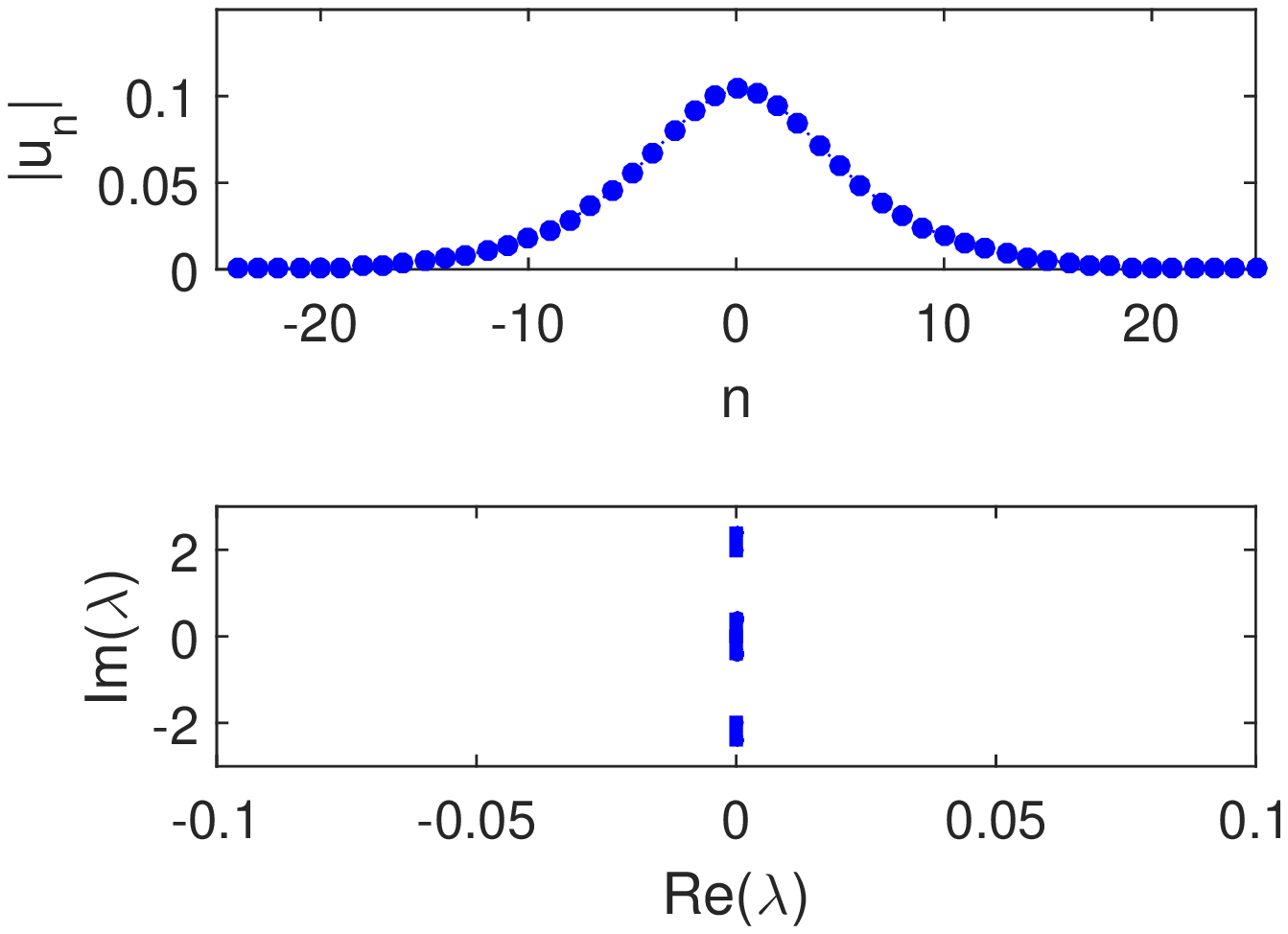}\label{subfig:prof_1_c_0_1_gamma_0_1}}
\subfigure[]{\includegraphics[scale=0.55]{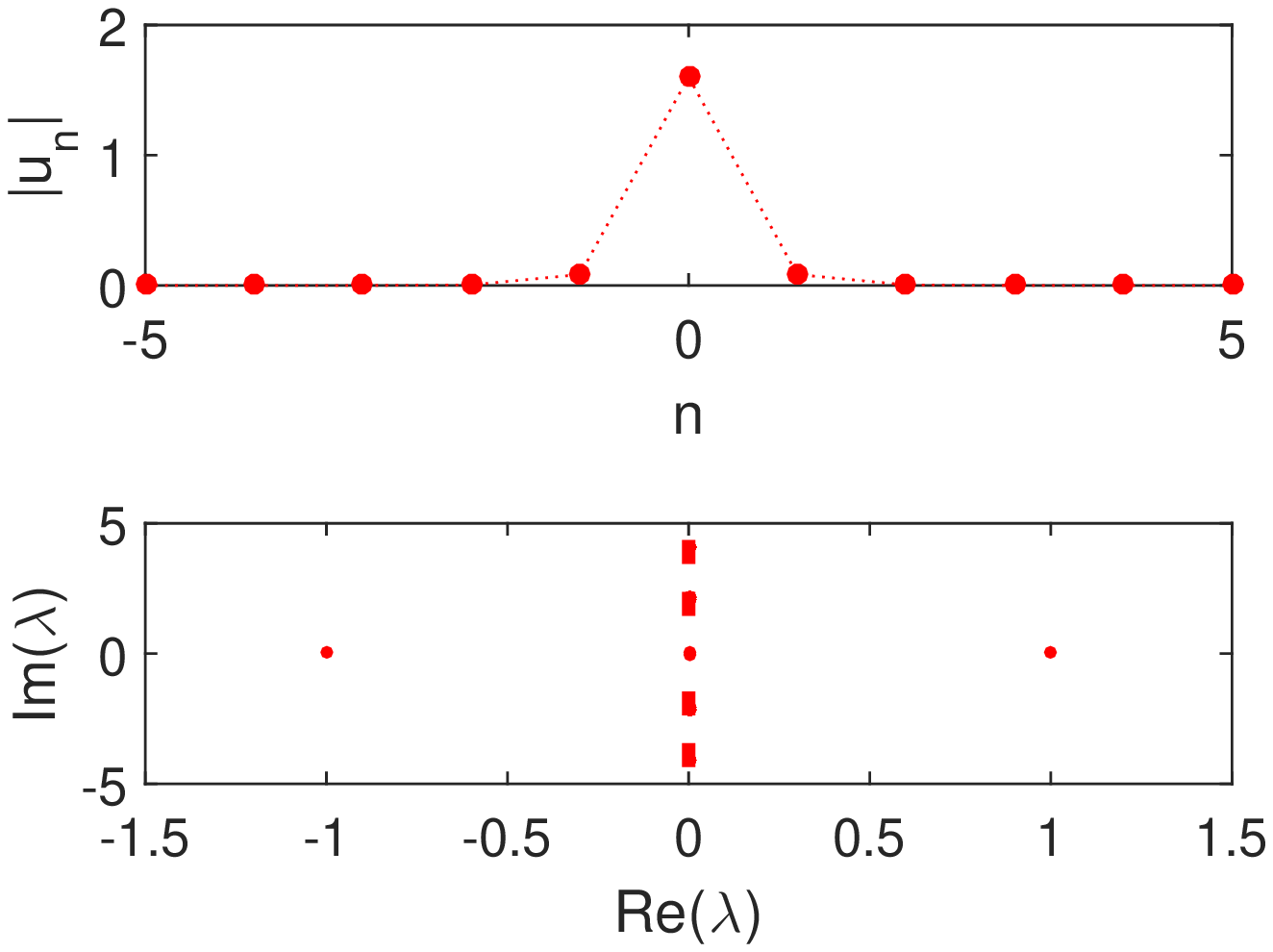}\label{subfig:prof_2_c_0_1_gamma_0_1}}
\subfigure[]{\includegraphics[scale=0.55]{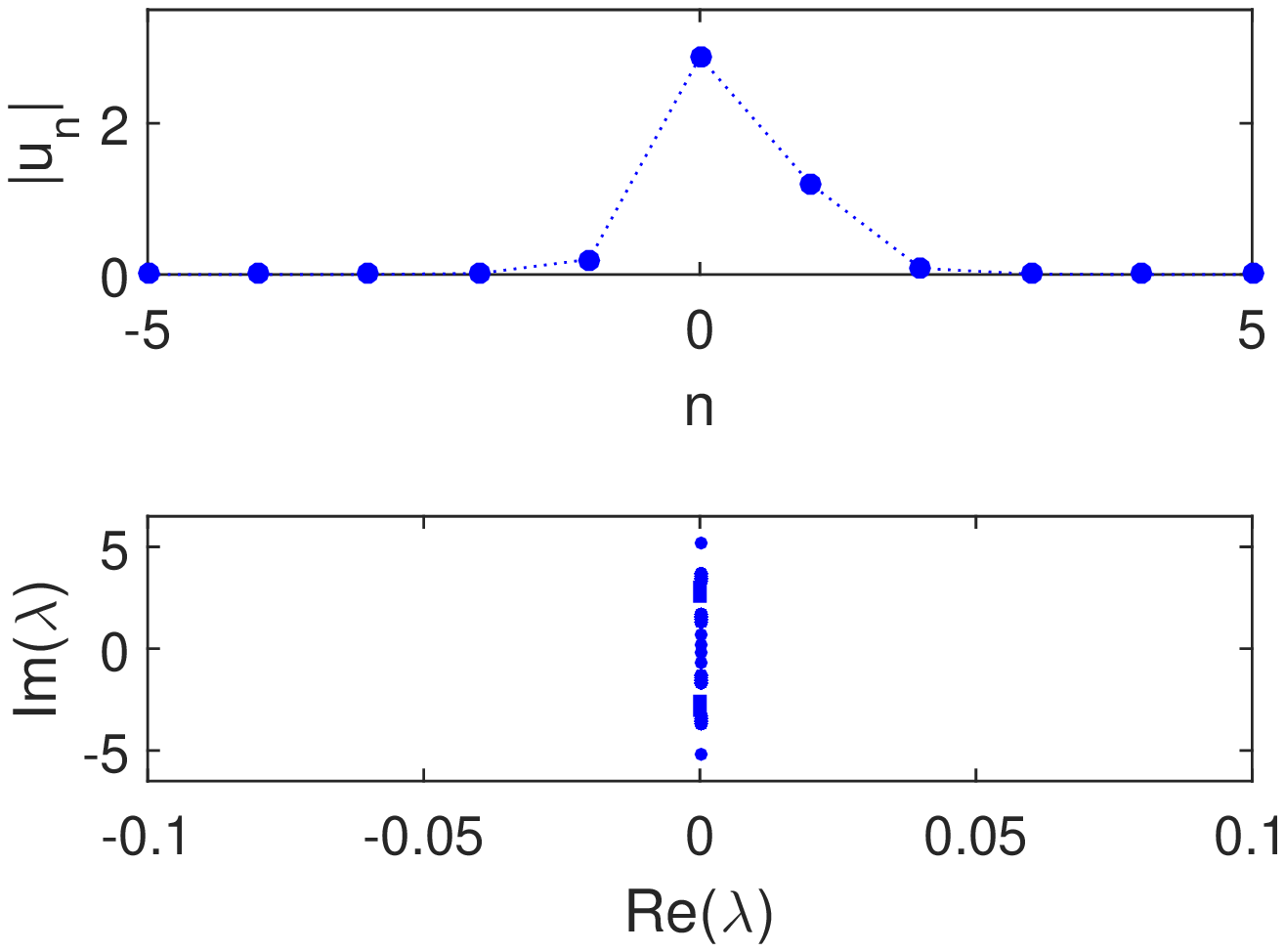}\label{subfig:prof_4_c_0_1_gamma_0_1}}
\subfigure[]{\includegraphics[scale=0.55]{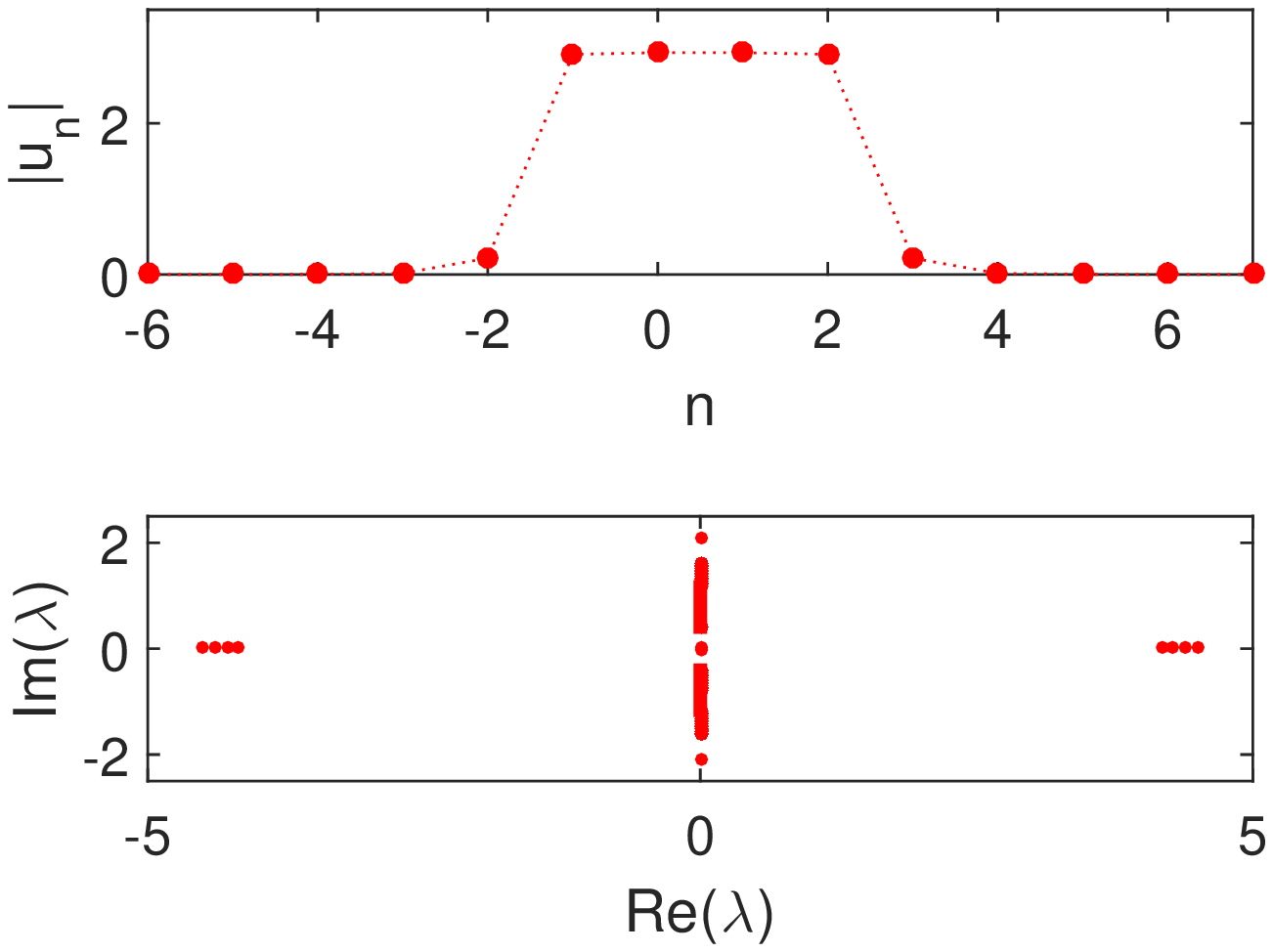}\label{subfig:prof_6_c_0_1_gamma_0_1}}
	\caption{Localised solutions on the bifurcation diagram shown in Fig.\ \ref{fig2} and their spectrum in the complex plane. Panels (a,b): bond-centred
		solutions. Panel (c): asymmetric solution, which has an intermediate shape between onsite and intersite profiles. Panel (d): site-centred solution.}
	\label{fig:prof}		
\end{figure*}

\begin{figure}[thbp!]
	\centering
	\subfigure[]{\includegraphics[scale=0.55]{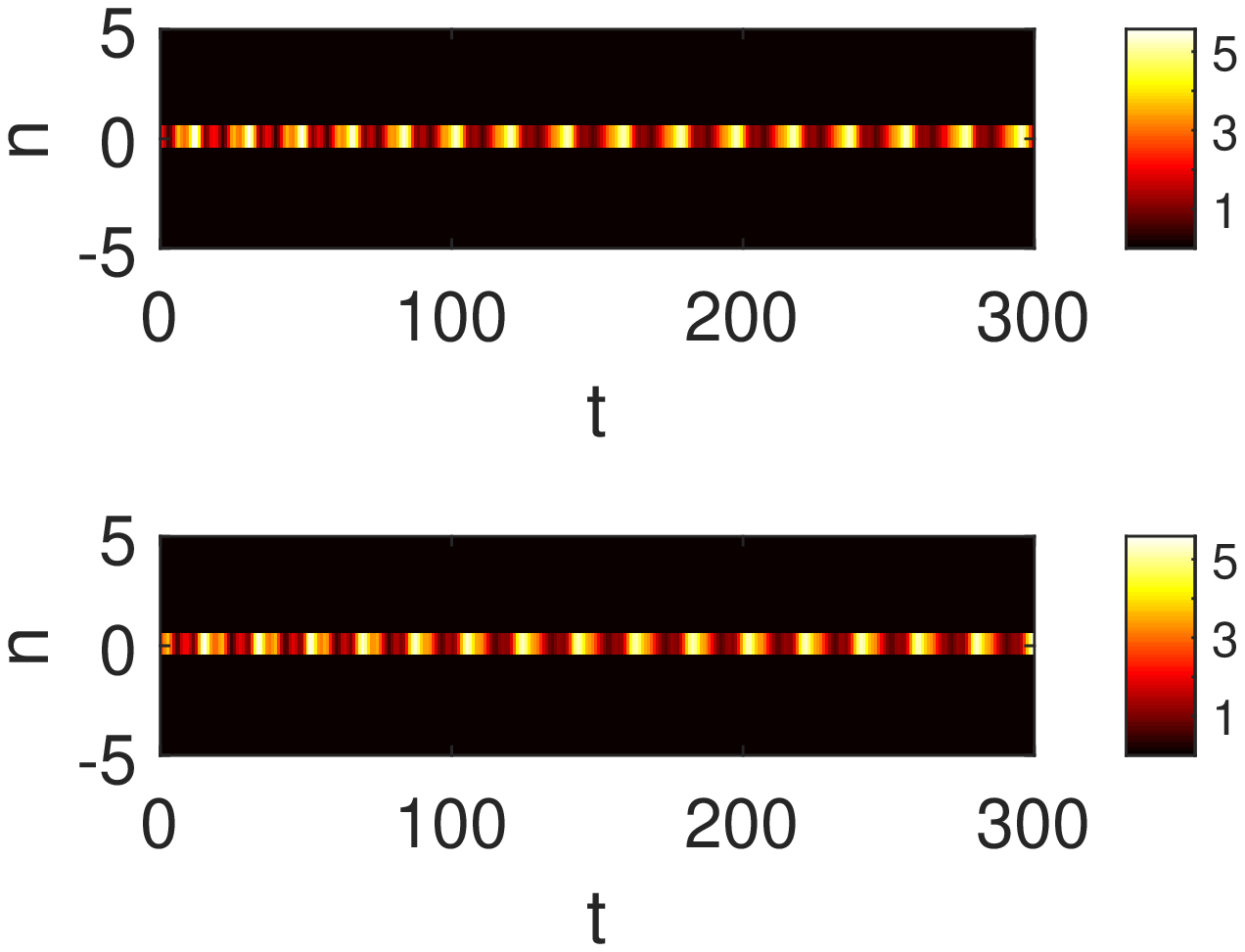}\label{subfig:topview_2_c_0_1_gamma_0_1}}
	\subfigure[]{\includegraphics[scale=0.55]{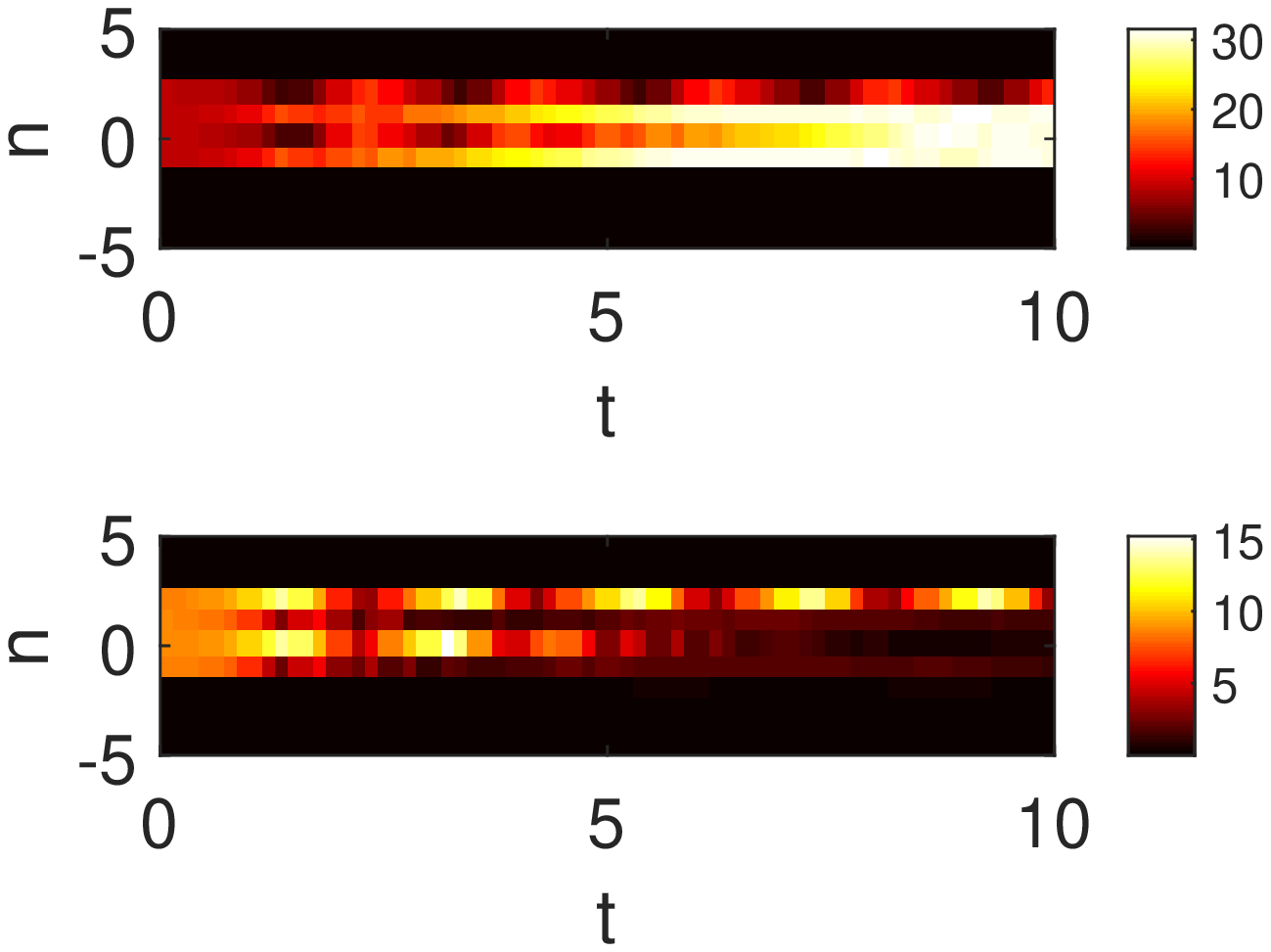}\label{subfig:topview_6_c_0_1_gamma_0_1}}
	\caption{Top view dynamics of the unstable solutions in Figs.\ \ref{subfig:prof_2_c_0_1_gamma_0_1} and \ref{subfig:prof_6_c_0_1_gamma_0_1}. Upper and lower panel is for $|u_n(t)|^2$ and $|v_n(t)|^2$, respectively.
	}
	\label{fig:dyn}		
\end{figure}

We consider discrete solitons of Eqs.\ \eqref{gov1} satisfying the localisation conditions $u_{n},v_{n}\rightarrow 0$ as $n\rightarrow \pm \infty$. It is known that there are two fundamental localized solutions 
existing for any coupling constant $C$, 
i.e.\ an intersite (bond-centred) and onsite (or site-centred) discrete mode with an even and odd number of high-intensity sites, respectively. 

Fixing the coupling $C$ and varying the propagation constant $\omega$, we depict the bifurcation diagrams of the two types of discrete modes in Fig.\ \ref{fig2}. For each  symmetric and antisymmetric configuration between $u_n$ and $v_n$, there are two branches that correspond to the site-centred and bond-centred solutions. 

In addition to symmetric solutions, there are also solutions that are asymmetric between the arms or asymmetric in the same arm. The former type of solutions corresponds to that giving 'as' branches in Fig.\ \ref{fig:uniform_ghost}, while the latter one constitutes the 'ladders' connecting snaking branches of onsite and intersite modes in Fig.\ \ref{fig2}. Both types emanate from pitchfork bifurcations (see, e.g., \cite{yang12} for relevant discussions on symmetry-breaking (pitchfork) bifurcations in generalized Schro\"odinger equations).

In Fig.\ \ref{fig:prof} we plot profiles of several localized solutions and their spectrum in the complex plane. Unstable solutions are due to spectra with nonzero real part, which belong to the red dashed segment in Fig.\ \ref{fig2}.

Bifurcation diagrams in Fig.\ \ref{fig2} form a snaking structure. Even though such structures have been reported before \cite{Carretero-Gonzalez2006,Chong2009,Chong2011,Taylor2010,Matthews2011}, the effect of the gain/loss parameter that yields different stability behaviours along the curves is novel. The region between the boundaries of the snakes is the pinning region. Comparing the two panels of Fig.\ \ref{fig2}, in agreement with the continuous case reported in \cite{burl13,burl16} the gain/loss parameter tends to destabilise localized solutions, shown by the dashed curve that tends to expand in the second panel. 

Up in the snaking structures (represented by, e.g., point 4 in Fig.\ \ref{fig2}(a)), the stability of the branches is similar to those in Fig.\ \ref{fig:uniform_ghost}. This is because the corresponding localized solutions have long plateau of nonzero uniform solutions, i.e.\ the stability is mainly determined by the continuous spectrum of the nonzero uniform solution.

We show in Fig.\ \ref{fig:dyn} the typical time evolution of unstable solutions in Fig.\ \ref{fig:prof}. While Fig.\ \ref{subfig:topview_2_c_0_1_gamma_0_1} indicates a clear blow up of the wave field with gain, which is common in \pts-systems \cite{pick13}, Fig.\ \ref{subfig:topview_6_c_0_1_gamma_0_1} shows intensity oscillations. The fact that the oscillations persist for quite a while is interesting by itself as \pts-symmetric dimers with cubic nonlinearity are known to have oscillations that blow up \cite{pick13}. Similar oscillations in the continuum limit $C\to\infty$ were also reported in \cite{burl16}, where the bounded oscillations were attributed to the quintic nonlinearity that may have suppressed the blow up. However, whether the long-live oscillation is a genuine cycle is addressed for future work. 

In the spatially uniform case, the branches of symmetric and antisymmetric solutions between the arms move towards each other as $\gamma$ increases and merge at $\gamma=1$. It is also the same with the case of localized solutions, i.e., the two snaking bifurcation diagrams in Fig.\ \ref{fig2} become closer with the increase of $\gamma$ and coincide at the critical value.

\section{Ghost states in the $\mathcal{PT}-$broken phase}
\label{sec3b}

\begin{figure}[thbp!]
	\centering
	{\includegraphics[scale=0.55]{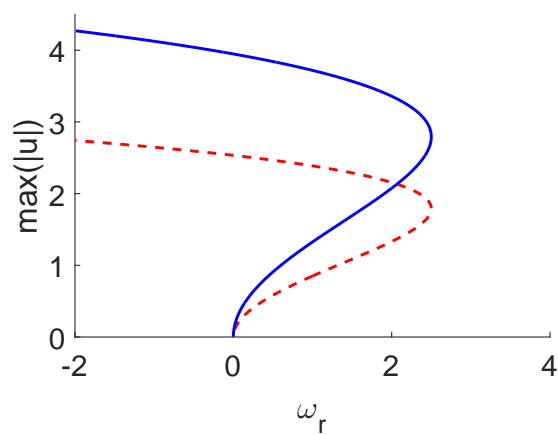}}
	\caption{Bifurcation diagram of spatially uniform ghost states in the broken $\mathcal{PT}-$symmetric region with {$\gamma=1.1$ and $C=0$}. 
	}
	\label{fig1}		
\end{figure}

In the broken $\mathcal{PT}-$symmetric region ($\gamma>1$), the trivial state $u_n=v_n=0$ is unstable. The typical time-evolution is that $u_n$ as the field with gain will blow-up, while $v_n$ that experiences loss decays. 

The \pts-phase transition ($\gamma=1$) is characterized by the merger of symmetric and antisymmetric solutions in a fold bifurcation. A follow-up question is what becomes of them past the critical point. It was also due to \cite{cart12} that it is possible to provide an analytic continuation for the original model in a nontrivial way. The continuation system is constructed by introducing a 'dual' system that 'forces' the solutions to mimic the $\mathcal{PT}-$symmetry of the potential in the original system, i.e., by setting $u_n^*\to v_n,\,v_n^*\to u_n$. In the broken $\mathcal{PT}$-symmetric phase $\gamma>1$, we therefore consider
\begin{equation}
\begin{aligned}
i\dot{u}_n&=\left(\mathcal{M} +i\gamma\right) u_n+v_n,\,
i\dot{v}_n=\left(\mathcal{M} -i\gamma\right) v_n+u_n,
\end{aligned}
\label{gov1gs}
\end{equation}
where $\mathcal{M}=C\Delta_{2}-\omega + u_nv_n -Q\left(u_nv_n\right)^2$. The parameter $\omega$ is again complex valued where the imaginary part must satisfy a self-consistency equation. Doing the same calculation, we also obtain Eq.\ \eqref{omi}. Because of complex $\omega$, Eqs.\ \eqref{gov1gs} are not \pts-symmetric and their solutions are also ghost states. The relation between \eqref{gov1} and \eqref{gov1gs} is that both systems yield the same symmetric and antisymmetric solutions at the phase transition $\gamma=1$.

We have computed the continuation of branches in Fig.\ \ref{fig:uniform_ghost} past the $\mathcal{PT}-$phase transition point. We present bifurcation diagrams of the ghost states in Fig.\ \ref{fig1}. We have also computed their stability from the corresponding linear eigenvalue problem of the dual system \eqref{gov1gs}. 

There are two uniform states that are mirror images of each other. Solutions with high intensity in $|u_n|$ are stable (in the sense of \eqref{gov1gs}), while the other ones with low $|u_n|$ are unstable, i.e., stable solutions correspond to $\sum_n|u_n|^2-|v_n|^2>0$. 

In the sense of the original system \eqref{gov1}, the stable solutions will lead to growth in time as the parameter $\omega_i$ is positive. On the other hand, the ones with negative $\omega_i$ decay in time and shall not be observed in direct numerical simulations. This thus means that ghost states of \eqref{gov1gs} may be interpreted as self-similar solutions of Eqs.\ \eqref{gov1}, at least for a short time \cite{achi14,li13}.

In Figs.\ \ref{fig:snake_ghost} and \ref{fig:ghost_prof} we plot bifurcation diagrams of localized ghost states 
and their profiles and stability computed through the 'dual' equations \eqref{gov1gs} and \eqref{omi}. We observe that the homoclinic snaking persists and that solutions with larger $|u_n|$ are stable (in the sense of the dual equations \eqref{gov1gs}). It is important to note that numerically the width of the pinning region of localized ghost states also does not depend on $\gamma$. 

\section{Analytical approximations}
\label{sec4}

\begin{figure}[htpb]
	\centering
	\subfigure[]{\includegraphics[scale=0.55]{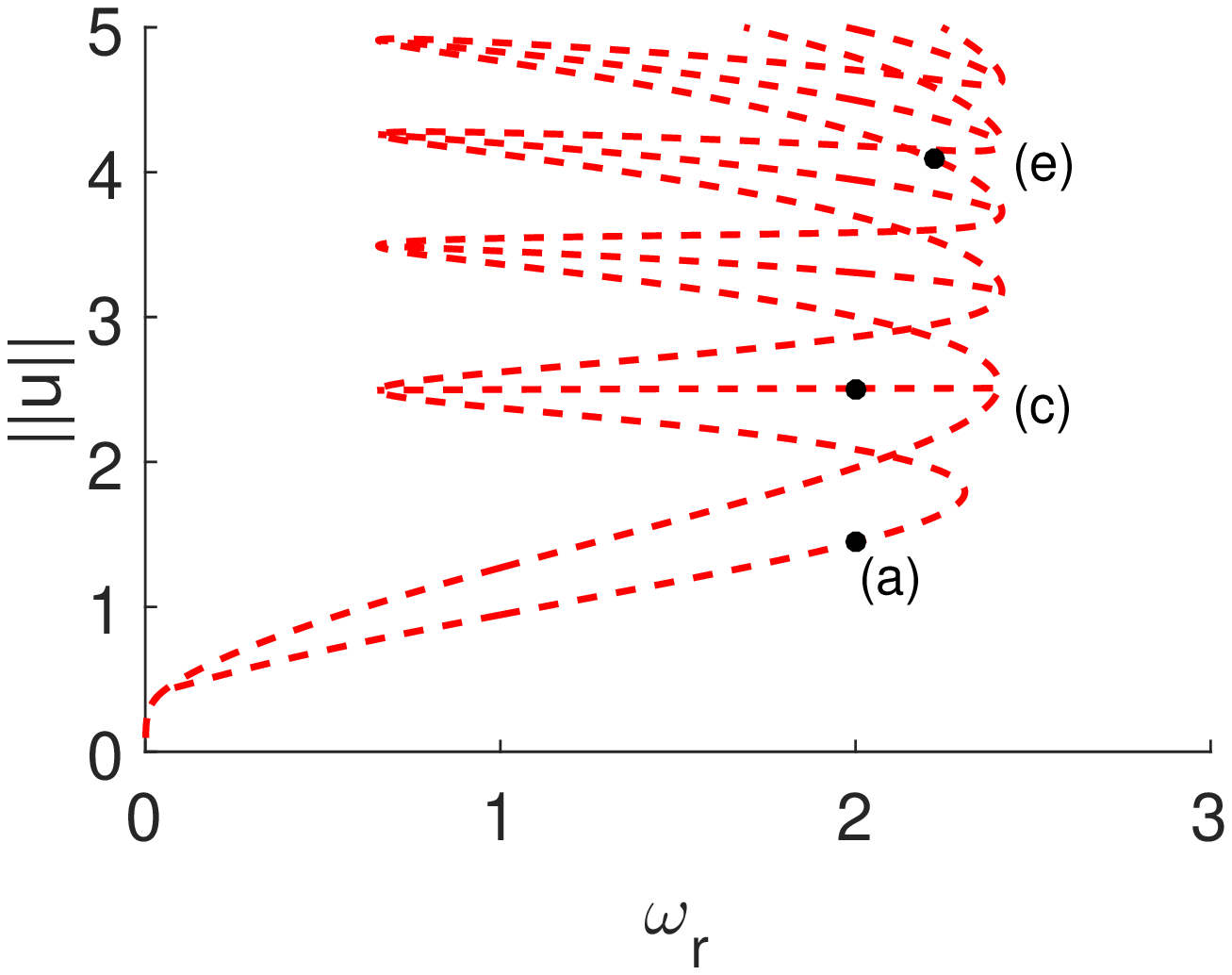}\label{subfig:bifur_c_0_1_gamma_1_1_snake1}}
	\subfigure[]{\includegraphics[scale=0.55]{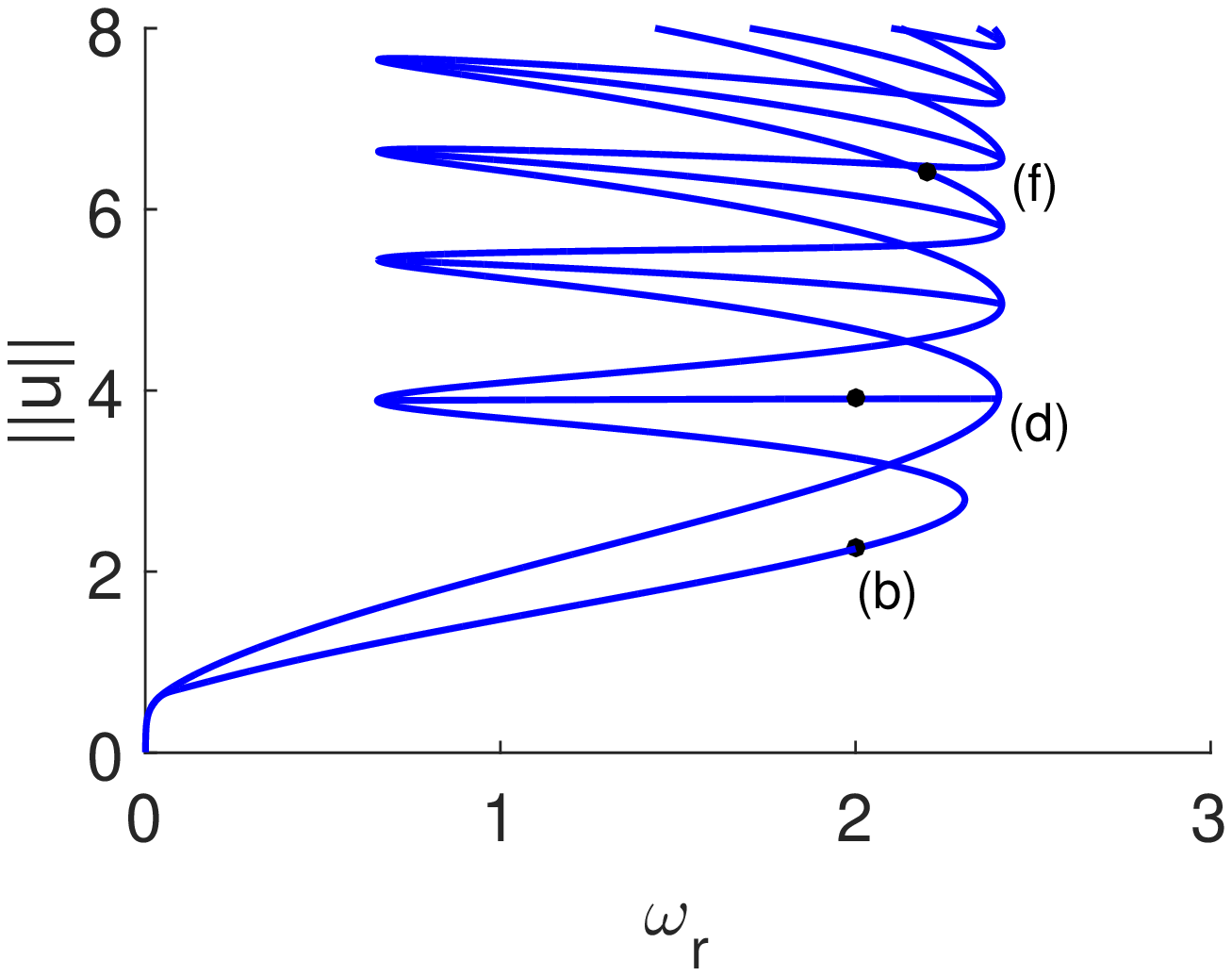}\label{subfig:bifur_c_0_1_gamma_1_1_snake2}}
	\caption{Bifurcation diagrams of localized solutions for the ghost states for {$\gamma=1.1$ and $C=0.1$}.}
	\label{fig:snake_ghost}
\end{figure}

\begin{figure*}[thbp!]
	\centering
	\subfigure[]{\includegraphics[scale=0.55]{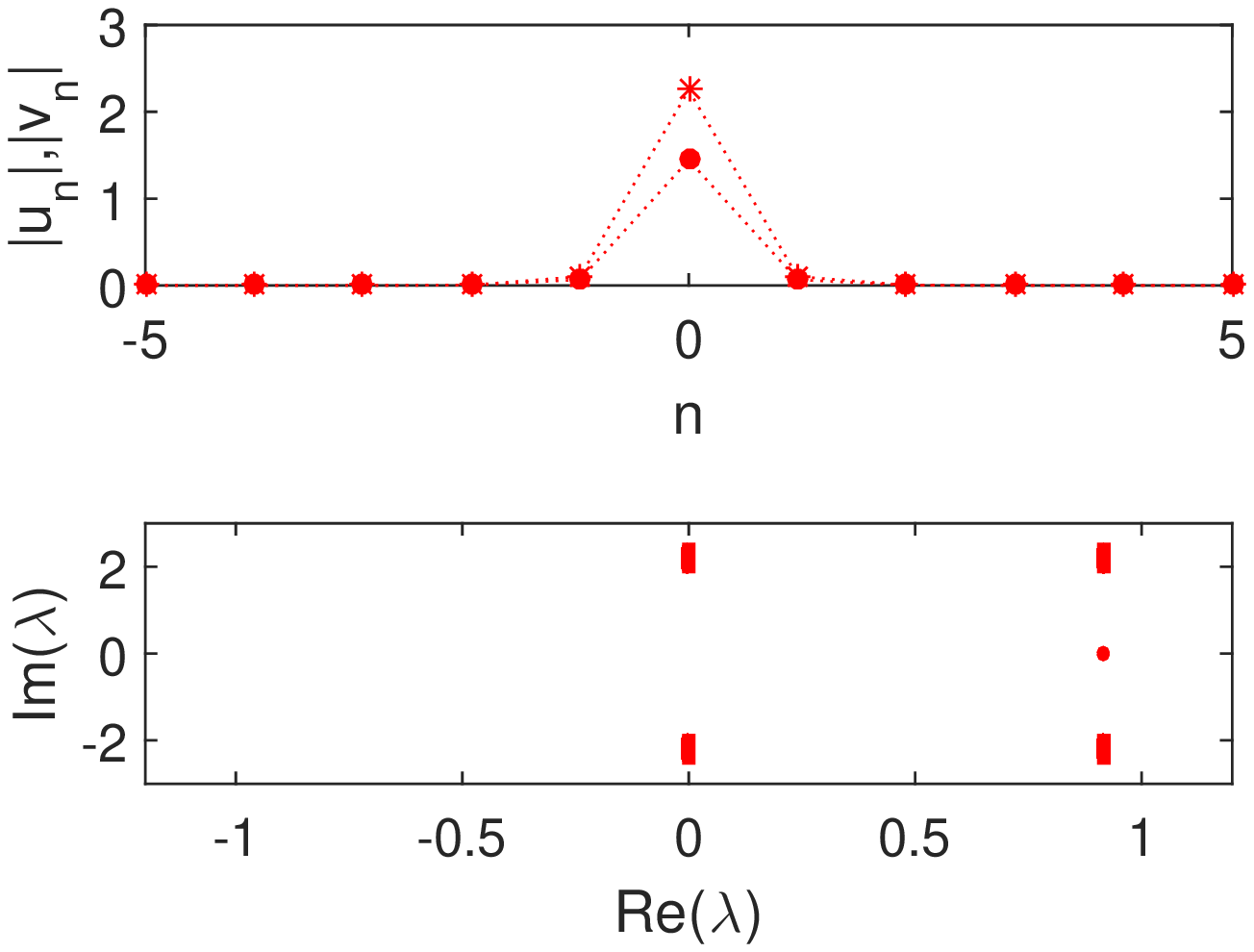}\label{subfig:prof_1_c_0_1_gamma_1_1_1}}
	\subfigure[]{\includegraphics[scale=0.55]{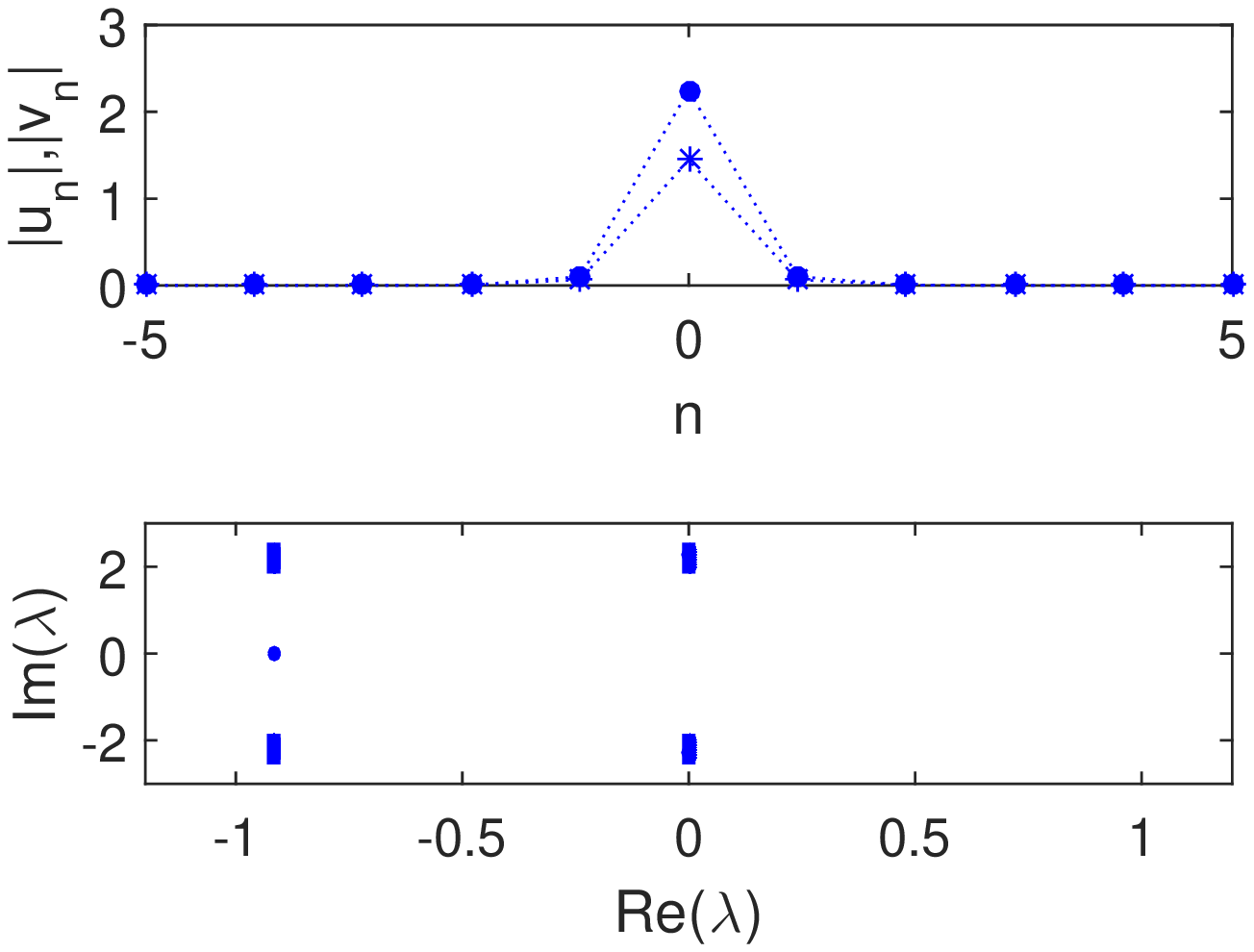}\label{subfig:prof_1_c_0_1_gamma_1_1_2}}
	\subfigure[]{\includegraphics[scale=0.55]{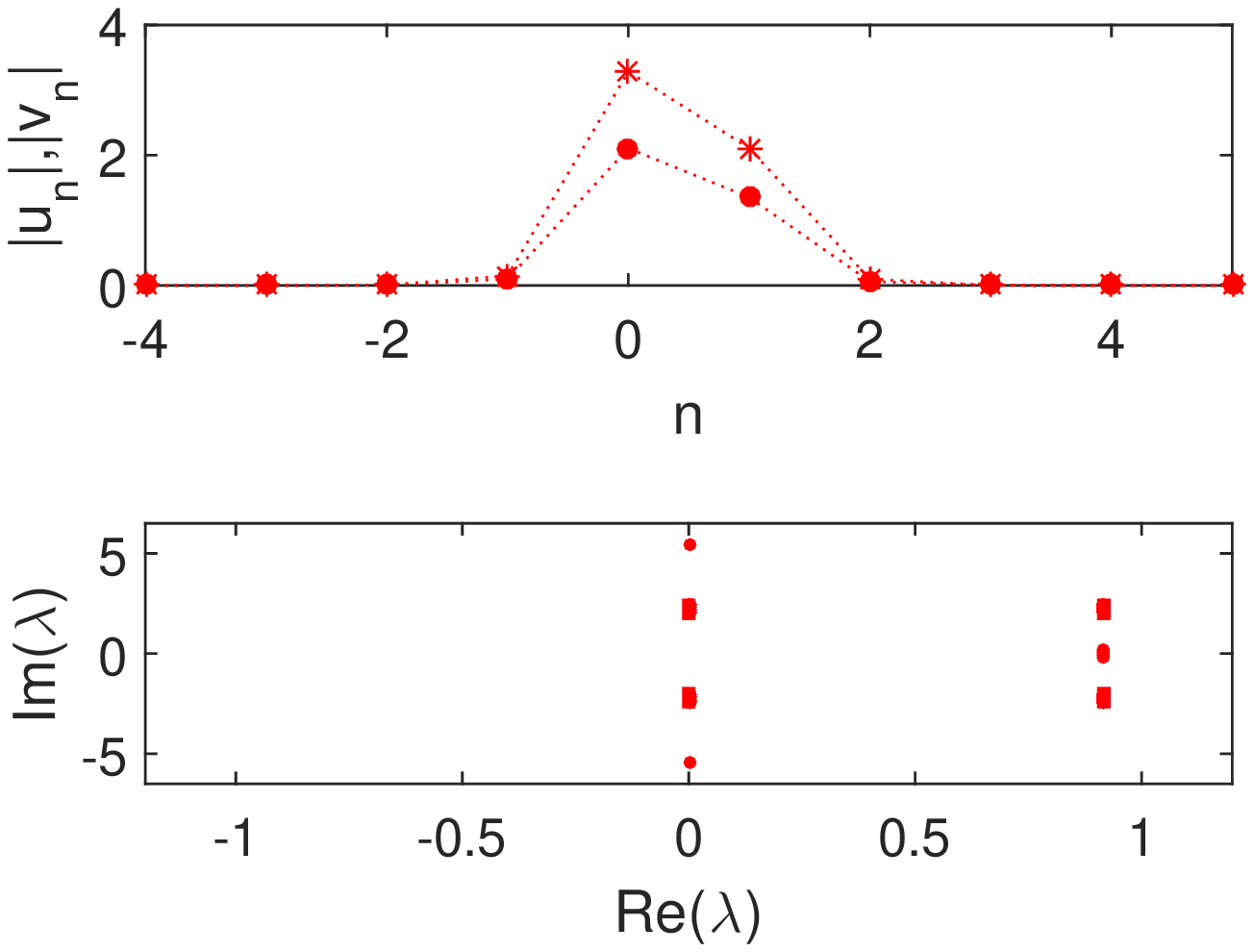}\label{subfig:prof_2_c_0_1_gamma_1_1_1}}
	\subfigure[]{\includegraphics[scale=0.55]{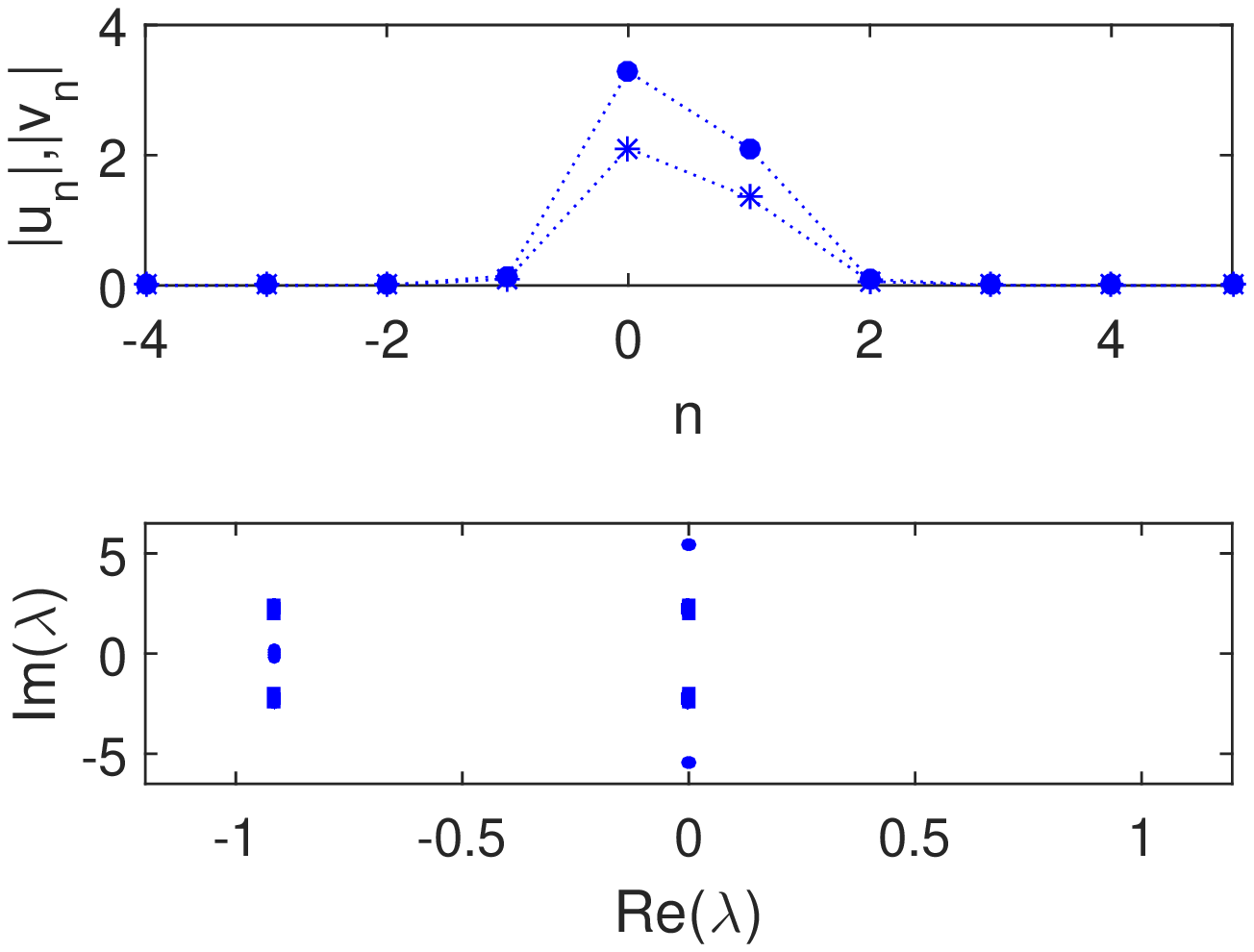}\label{subfig:prof_2_c_0_1_gamma_1_1_2}}
	\subfigure[]{\includegraphics[scale=0.55]{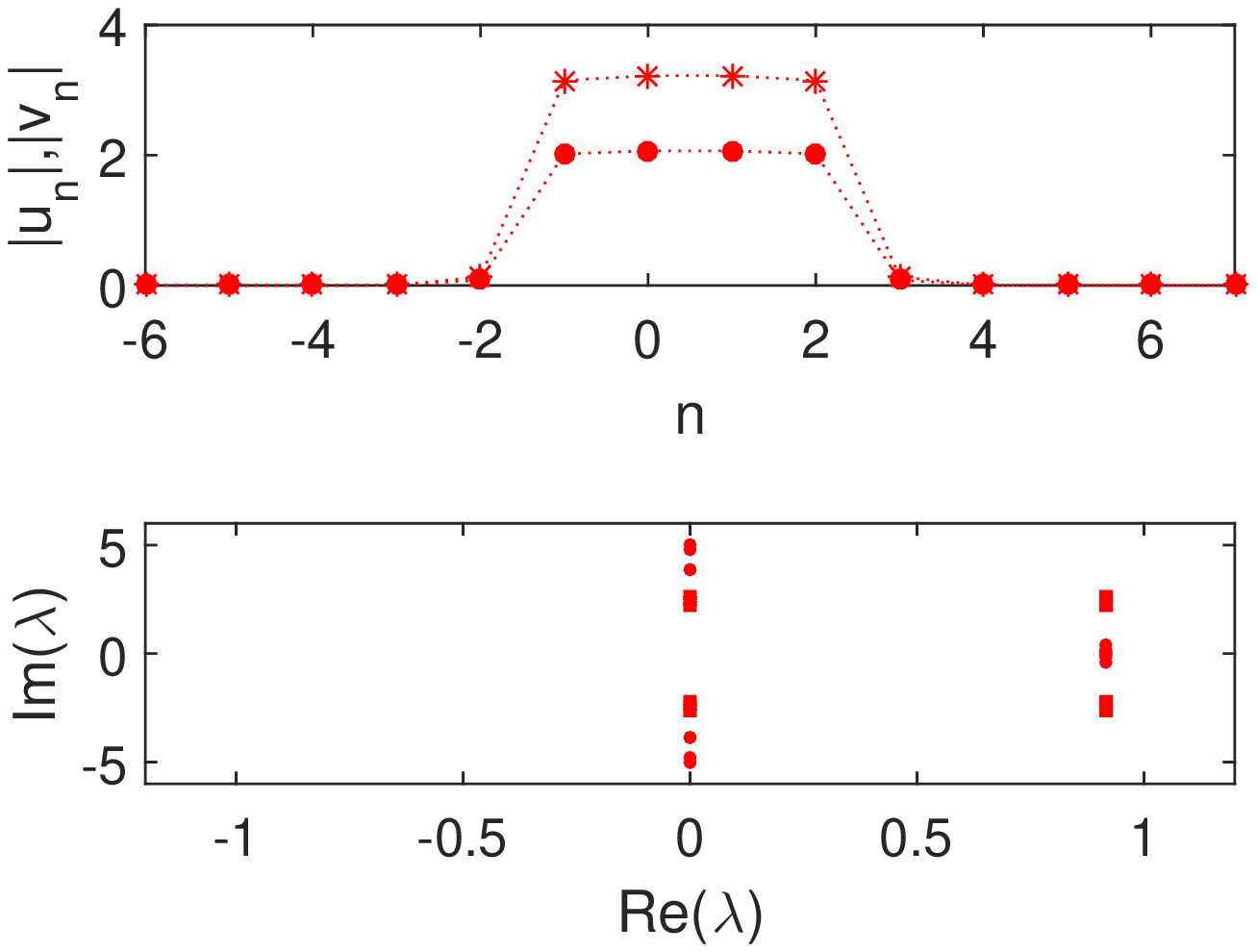}\label{subfig:prof_3_c_0_1_gamma_1_1_1}}
	\subfigure[]{\includegraphics[scale=0.55]{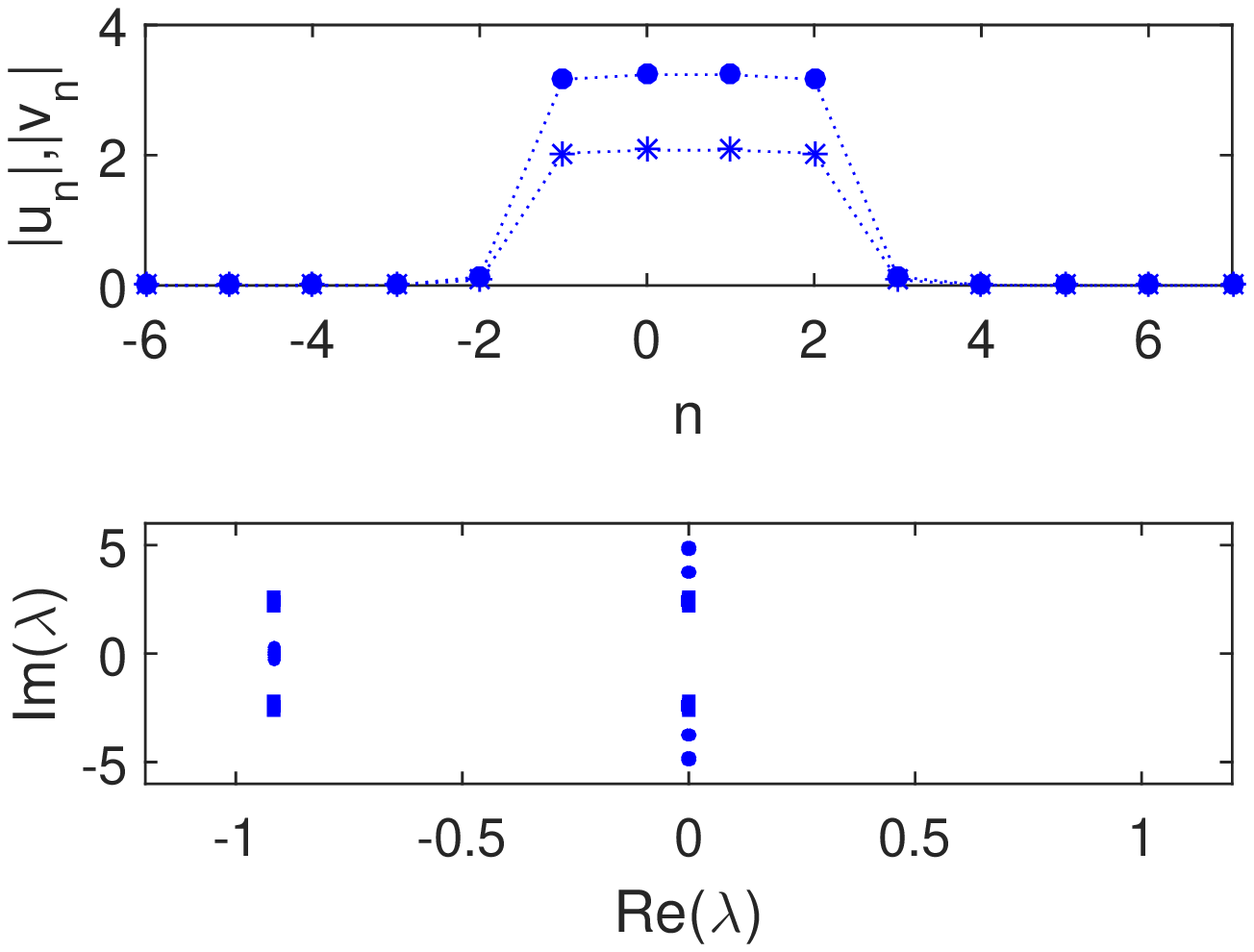}\label{subfig:prof_3_c_0_1_gamma_1_1_2}}
	\caption{Plot of the localized ghost states on the bifurcation diagram shown in Fig.\ \ref{fig:snake_ghost} and their spectrum in the complex plane. Panels (a,c,e): unstable
		solutions. Panels (b,d,f): stable solutions. 
		$|u_n|$ and $|v_n|$ are represented by circle and star points, repectively.
	}
	\label{fig:ghost_prof}		
\end{figure*}

In this section, we will study the width of the snaking region in Fig.\ \ref{fig2} as a function of, e.g., the coupling constant $C$. We will derive an asymptotic approximation of the width. The approach is distinguished in two different regions, i.e.\ small and large coupling. Because the width of the pinning region is independent of $\gamma$, our result is also applicable for the snaking region of localized ghost states in Fig.\ \ref{fig:snake_ghost}.


\subsection{Small coupling case}

When $C$ is small, as we follow the snaking structure upward (see Fig.\ \ref{fig2}), at the leading order there is only one site that is `active', with the remaining sites being either at $0$ or at the plateau of a nonzero uniform solution. Such behaviours were observed and exploited in many ways before, see, e.g., \cite{carp03,wolf12}, but not in the context of snaking.

From \eqref{usol}, we assume that up in the snaking diagram only the following nodes are involved in the dynamics, i.e.\
\begin{eqnarray}
u_{n-1}=0,\quad  u_n=a, \quad u_{n+2}=\sqrt{\frac{1+\sqrt{1-4Q\Omega}}{2Q}}.
\label{eq:coupling_sh_assume}
\end{eqnarray}
Note that we only use the `$+$' sign for the uniform solution forming the plateau, which is the upper branch in Fig.\ \ref{fig:uniform_ghost}. Substituting \eqref{eq:coupling_sh_assume} into the time-independent discrete equation \eqref{gov3} will yield the one-active-site approximation: 
\begin{equation}
f(a):=Q a^5 - a^3 +(\Omega+2C) a - C\sqrt{\frac{1+\sqrt{1-4Q\Omega}}{2Q}}=0.
\label{1site}
\end{equation}
. 

In general \eqref{1site} will have five roots. The roots relevant to our study are the positive ones. As $\Omega$ varies, two of the roots will collide in a saddle-centre bifurcation. This condition corresponds to the boundaries of the snaking region. The condition for the collision is when a local maximum or minimum of the function $f(a)$ crosses the horizontal axis. The critical points of $f(a)$ are given by
\begin{equation}
f'(a):=5Q a^4 - 3a^2 +(\Omega+2C)=0,
\label{1sited}
\end{equation}
i.e.\
\begin{equation}
a=\sqrt{\frac{3\pm\sqrt{9-20Q(\Omega+2C)}}{10Q}}.
\label{ac}
\end{equation}

Substituting \eqref{ac} into \eqref{1site} and solving the resulting equation for $\Omega$ asymptotically give us
\begin{equation}
\Omega=\frac{1}{4Q}-C+\mathcal{O}(C^2),\, 3\sqrt[3]{\frac{C^2}{{4Q}}}+\mathcal{O}(C^{4/3}).
\label{cs}
\end{equation}
The snaking width $W$ is then given approximately by the difference between the two functions.

\subsection{Large coupling case}

Following \cite{flac96,sagd88}, Eq.\ \eqref{gov3} is identical to the equation
\begin{equation}
C\tilde{A}_{xx}+\sum_{n=-\infty}^\infty\delta(x-n) F(\tilde{A}(x))=0,
\label{gov4}
\end{equation}
where $A_n=\tilde{A}(x=n)$ and $F(\tilde{A})=-\Omega \tilde{A} +\tilde{A}^3-Q\tilde{A}^5$. The proof is as follows. 

First, from \eqref{gov4}, we obtain that $\tilde{A}_{xx}(n<x<n+1)=0$ or upon integration $\tilde{A}_{x}(n<x<n+1)=\text{const}$. Thus, 
\begin{equation}
\tilde{A}_x(n+1/2)=\tilde{A}(n+1)-\tilde{A}(n).
\label{s4}
\end{equation}

Next, integrate \eqref{gov4} from $x=n-1/2$ to $x=n+1/2$ to obtain
\begin{equation}
C\left(\tilde{A}_{x}(n+1/2)-\tilde{A}_{x}(n-1/2)\right)=F(\tilde{A}(n)).
\label{s5}
\end{equation}
Using Eq.\ \eqref{s4}, Eq.\ \eqref{s5} becomes
the lattice equation \eqref{gov3}.

Using Fourier series, we can then write the summation $\sum_{n=-\infty}^\infty\delta(x-n)=1+2\sum_{k=1}^\infty\cos(2\pi kx)$, which converges to the Dirac comb non-uniformly. Taking only the first harmonic, \eqref{gov4} then becomes
\begin{equation}
C\tilde{A}_{xx}+(1+2\cos(2\pi x))\left(-\Omega \tilde{A} +\tilde{A}^3-Q\tilde{A}^5\right)=0,
\label{govt}
\end{equation}
which can be expected to approximate \eqref{gov3} in the large coupling case for $C\gg1$ \cite{flac96}.

Without the periodic potential $2\cos(2\pi x)$, Eq.\ \eqref{govt} 
has a front solution given by
\begin{equation}
\tilde{A}(x)=\sqrt{\frac{3}{{4}{Q\left(1+e^X\right)}}},\, X={\sqrt{\frac3{4QC}}{x}},
\label{fron}
\end{equation}
when \begin{equation}
\Omega={3}/(16Q).
\label{mp}
\end{equation}

Following \cite{Susanto2011,Matthews2011}, we will approximate the solutions along the snaking structure by
\begin{equation}
\tilde{A}(x)=\sqrt{3}/\sqrt{{{4}{Q\left(1+e^{{(|X-\phi|-LX/x)}}\right)}}},
\label{fr}
\end{equation}
where $\phi$ is the phase-shift distinguishing the two branches, i.e.\ $\phi=0,\,X/(2x)$ for the on-site and intersite solutions, respectively. $L$ is the length of the plateau, which is presently an unknown variable.

Using the standard variational argument, requiring \eqref{fr} to be an optimal solution of \eqref{govt} implies that $L$ must satisfy the equation (see, e.g., \cite{dawe13})
\begin{equation}
\int_{-\infty}^\infty \left[C\tilde{A}_{xx}+(1+2\cos(2\pi x))F(\tilde{A})\right] \frac{\partial \tilde{A}}{\partial L}\,dx=0,
\label{solvb}
\end{equation}
where $\Omega$ is set to be near the Maxwell point \eqref{mp}, i.e.\ $\Omega=3/(16Q)+\Delta\Omega$.

Equation \eqref{solvb} can be simplified at the leading order for $L\gg1$ to
\begin{align}
\displaystyle
\Delta\Omega&=\lim_{L\to\infty} \frac{-2C\int_{0}^\infty \cos(2\pi x)\tilde{A}_{xx}\frac{\partial \tilde{A}}{\partial L}\,dx}{\int_{0}^\infty \tilde{A}\frac{\partial \tilde{A}}{\partial L}\,dx}\nonumber\\
&=\frac{C\pi^3}3\csch\left(\frac{4\pi^2\sqrt{CQ}}{\sqrt{3}}\right)\nonumber\\
&\times\left[4\pi\sqrt{3CQ}\cos({2\pi L})+3\sin(2\pi L)\right].
\label{solvb1}
\end{align}
The width of the snaking region is then simply given by
\begin{align}
W &=  \frac{2C\pi^3}3\csch\left(\frac{4\pi^2\sqrt{CQ}}{\sqrt{3}}\right)\sqrt{48\pi^2CQ+9},\label{exp1}\\
&\approx  {16\pi^4C^{3/2}}\sqrt{\frac{Q}{3}}e^{-{4\pi^2\sqrt{CQ/3}}},\label{exp2}
\end{align}
which is exponentially small. 

\begin{figure}[tbhp]
\subfigure[]{\centerline{\includegraphics[width=8cm]{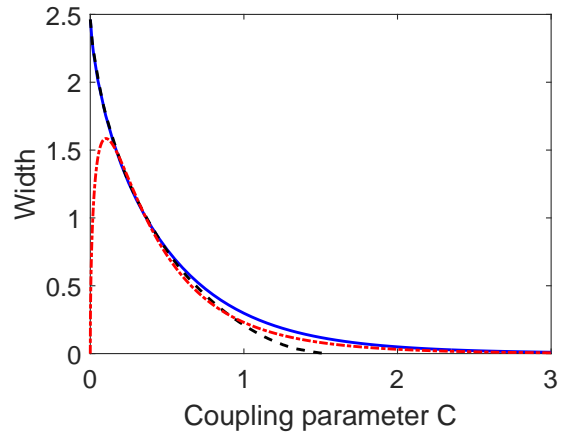}}}
\subfigure[]{\centerline{\includegraphics[width=8cm]{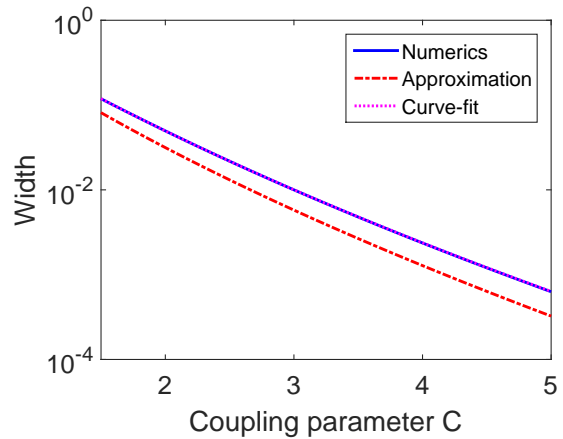}}}
	\caption{The width of the snaking region as a function of the coupling constant $C$ for $Q=0.1$. The solid curve is obtained from the numeric and the dashed and dash-dotted lines are the approximations \eqref{cs} for $0\leq C\ll1$ and \eqref{exp1} for $C\gg1$, respectively. In panel (b), we also plot a curve fit \eqref{exp3}, that is indistinguishable from the numerical curve.}
	\label{fig5}
\end{figure}


As pointed out by one of the referees, the exponential factor in the approximation  \eqref{exp1}-\eqref{exp2} is correct, which can be justified in the following way, explained in details in \cite{kozy13}. The continuous problem \eqref{govt} involves two scales: a fast scale variable, $x$, and a slow one, $X$. These two scales are infinitely separated in the limit $C\to\infty$, i.e., $X/x\to0$.  The front solution \eqref{fron} contains singularities in the complex plane, the closest to the real axis being $X_0 = i\pi$, which then leads to the exponential part of the pinning. This was used in \cite{kozy13} to derive the dependance of the pinning range on the front orientation in general 2D pattern forming systems. As for the algebraic factor, $C^{3/2}$, only a proper exponential asymptotic treatment (see, e.g., \cite{king01} for exponential asymptotic analysis for a nonlinear differential difference equation similar to \eqref{gov1}) can establish its exponent. The exponential and algebraic scales in \eqref{exp2} are, however, in agreement with those obtained using exponential asymptotics in \cite{Dean2015} for homoclinic snaking on a planar lattice, i.e., of two-dimensional fronts that are localized in one direction only.

We show in Fig.\ \ref{fig5}(a) the width of the snaking region computed numerically and our approximations \eqref{cs} and \eqref{exp1}. One can see good agreement between them.

Note that Fig.\ \ref{fig5}(a) does not allow one to check \eqref{exp1} because of the very fast exponential decay as $C$ increases. We depict in Fig.\ \ref{fig5}(b) the comparison in a log plot, where it is clear that the approximation deviates from the numeric as $C$ increases. Using the function
\begin{align}
	W =  \alpha C^\beta e^{-{4\pi^2\sqrt{CQ/3}}},\label{exp3}
\end{align}
we curve fit the numerical result where obtain that $\alpha=405.03$ and $\beta=1.71$. There is a slight difference in the algebraic scale, which may be attributed to the step of taking the first harmonic only in Eqs.\ \eqref{govt}.

\section{Conclusion}
\label{sec5}

Spatially uniform and localized solutions (site-centered and bond-centered modes) and their bifurcation diagrams that form a snaking structure in a parity-time ($\cal{PT}$)-symmetric coupler composed by a chain of dimers have been discussed. It has been shown that the gain/loss coefficient does not influence the width of the snakes. 

In the broken $\mathcal{PT}-$symmetry region $\gamma>1$, we have also analysed the continuations of the time-independent solutions, that are called ghost states. Interestingly localized ghost states have also been observed to exhibit a homoclinic snaking in their bifurcation diagrams, with the same width of pinning region as that of the localized solutions with $0<\gamma<1$. 

Asymptotic approximations of the width of the snaking region have been derived in two different limits, i.e.\ strong and weak coupling between the dimers. The approximations have been compared with numerical results where good agreement is obtained.


	\section*{Acknowledgement}
	
	We acknowledge the two referees for their detailed and valuable suggestions. The H.S.\ and N.L.\ acknowledge financial support from the UK Engineering and Physical Sciences Research Council (Grant No. EP/M024237/1). R.K.\ gratefully acknowledges financial support from Lembaga Pengelolaan Dana Pendidikan (Indonesia Endowment Fund for Education) Grant No.- Ref: S-34/LPDP.3/2017. H.S., D.A., E.R.M.P., and T.A.\ are grateful to the Ministry of Research, Technology and Higher Education of the Republic of Indonesia for the World Class Professor program, contract number 168.A15/D2/KP/2017, that led to this joint publication. 
	
	H.S.\ and R.K.\ contributed equally to this work.
	


\begin{thebibliography}{999}

\bibitem{Purwins2010} H.-G. Purwins, H. U. Bo\"edeker, and Sh. Amiranashvili, \href{http://dx.doi.org/10.1080/00018732.2010.498228}{Adv. Phys. \textbf{59}, 485 (2010)}.
\bibitem{Barbay2008} S. Barbay, X. Hachair, T. Elsass, I. Sagnes, and R. Kuszelewicz, \href{https://doi.org/10.1103/PhysRevLett.101.253902}{Phys. Rev. Lett. \textbf{101}, 253902 (2008)}.
\bibitem{Haudin2011} F. Haudin, R. G. Rojas, U. Bortolozzo, S. Residori, and M. G. Clerc, \href{https://doi.org/10.1103/PhysRevLett.107.264101}{Phys. Rev. Lett. \textbf{107}, 264101 (2011)}.
\bibitem{Lloyd2015} D. J. B. Lloyd, C. Gollwitzer, I. Rehberg, and R. Richter,	\href{https://doi.org/10.1017/jfm.2015.565}{J. Fluid Mech. \textbf{783}, 283 (2015)}.

\bibitem{coul00} P.\ Coullet, C.\ Riera, and C.\ Tresser, 
Phys. Rev. Lett. 84, 3069-3072 (2000).

		
\bibitem{Woods2006} P. D. Woods and A. R. Champneys, \href{http://dx.doi.org/10.1016/S0167-2789(98)00309-1}{Physica D \textbf{129}, 147 (1999)}.

\bibitem{Pomeau1986} Y. Pomeau, \href{http://dx.doi.org/10.1016/0167-2789(86)90104-1}{Physica D \textbf{23}, 3 (1986)}.
\bibitem{Burke2006} J. Burke, E. Knobloch, \href{https://doi.org/10.1103/PhysRevE.73.056211}{Phys. Rev. E \textbf{73}, 056211 (2006)}.
\bibitem{Burke2007} J. Burke and E. Knobloch, \href{http://dx.doi.org/10.1063/1.2746816}{Chaos \textbf{17}, 037102 (2007)}.
\bibitem{Burke2007a} J. Burke and E. Knobloch, \href{Phys. Lett. A 360, 681 (2007)}{Phys. Lett. A \textbf{360}, 681 (2007)}.
\bibitem{Sakaguchi1996} H. Sakaguchi and H. R. Brand, \href{http://dx.doi.org/10.1016/0167-2789(96)00077-2}{Physica D \textbf{97}, 274 (1996)}.
\bibitem{Chapman2009} S. J. Chapman and G. Kozyreff, \href{http://dx.doi.org/10.1016/j.physd.2008.10.005}{Physica D \textbf{238}, 319 (2009)}.
\bibitem{Kozyreff2006} G. Kozyreff and S. J. Chapman, \href{https://doi.org/10.1103/PhysRevLett.97.044502}{Phys. Rev. Lett. \textbf{97}, 044502 (2006)}.

\bibitem{Dean2011} A. D. Dean, P. C. Matthews, S. M. Cox, and J. R. King. \href{http://dx.doi.org/10.1088/0951-7715/24/12/003}{Nonlinearity \textbf{24}, 3323 (2011)}.
\bibitem{Susanto2011} H. Susanto and P. C. Matthews, \href{https://doi.org/10.1103/PhysRevE.83.035201}{Phys. Rev. E \textbf{83}, 035201(R) (2011)}.
\bibitem{Carretero-Gonzalez2006} R. Carretero-Gonzalez, J. D. Talley, C. Chong, and B. A. Malomed, \href{http://dx.doi.org/10.1016/j.physd.2006.01.022}{Physica D \textbf{216}, 77 (2006)}.
\bibitem{Chong2009} C. Chong, R. Carretero-Gonzalez, B. A. Malomed and P. G. Kevrekidis, \href{http://dx.doi.org/10.1016/j.physd.2008.10.002}{Physica D \textbf{238}, 126 (2009)}.
\bibitem{Chong2011} C. Chong and D. E. Pelinovsky, \href{http://dx.doi.org/10.3934/dcdss.2011.4.1019}{Disc. Cont. Dyn. Sys. S \textbf{4}, 1019 (2011)}.
\bibitem{Taylor2010} C. Taylor and J. H. P. Dawes, \href{http://dx.doi.org/10.1016/j.physleta.2010.10.010}{Phys. Lett. A \textbf{375}, 14 (2010)}.
\bibitem{Yulin2008} A. V. Yulin, A. R. Champneys, and D. V. Skryabin, \href{https://doi.org/10.1103/PhysRevA.78.011804}{Phys. Rev. A \textbf{78}, 011804(R) (2008)}.
\bibitem{Yulin2010} A. V. Yulin and A. R. Champneys, \href{http://dx.doi.org/10.1137/080734297}{SIAM J. Appl. Dyn. Syst. \textbf{9}(2), 391 (2010)}.
\bibitem{Clerc2011} M. G. Clerc, R .G. Elias and R. G. Rojas, \href{https://doi.org/10.1098/rsta.2010.0255}{Phil. Trans. Roy. Soc. A \textbf{369}, 412 (2011)}.
\bibitem{mccu16} N.\ McCullen and T.\ Wagenknecht, Scientific Reports 6, 27397 (2016).
\bibitem{Matthews2011} P. C. Matthews and H. Susanto, \href{https://doi.org/10.1103/PhysRevE.84.066207}{Phys. Rev. E \textbf{84}, 066207 (2011)}.
\bibitem{Dean2015} A. D. Dean, P. C. Matthews, S. M. Cox, and J. R. King, \href{http://dx.doi.org/10.1137/140966897}{SIAM J. Appl. Dyn. Syst. \textbf{14}(1), 481 (2015)}.
\bibitem{such16} Suchkov, S. V., A. A. Sukhorukov, J. Huang, S. V. Dmitriev,
C. Lee, Yu. S. Kivshar, 2016, Laser \& Photon.\ Rev.\ 10, 177.

\bibitem{kono16} V.\ V.\ Konotop, J. Yang, and D. A. Zezyulin, Rev. Mod. Phys. 88, 035002 (2016).





\bibitem{bend98} C. M. Bender and S. Boettcher, \href{https://doi.org/10.1103/PhysRevLett.80.5243}{Phys. Rev. Lett. 80, 5243 (1998)}.
\bibitem{bend99} C. M. Bender, S. Boettcher, and P. N. Meisinger, \href{http://dx.doi.org/10.1063/1.532860} {J. Math. Phys. 40, 2201 (1999)}.
\bibitem{bend07} C. M. Bender, \href{https://doi.org/10.1088/0034-4885/70/6/R03}{Rep. Prog. Phys. 70, 947 (2007)}.
\bibitem{mois11} N. Moiseyev, \emph{Non-Hermitian Quantum Mechanics} (Cambridge University Press, Cambridge, 2011).

\bibitem{guo09} A. Guo, G. J. Salamo, D. Duchesne, R. Morandotti, M. Volatier-Ravat, V. Aimez, G. A. Siviloglou, and D. N. Christodoulides, \href{https://doi.org/10.1103/PhysRevLett.103.093902}{Phys. Rev. Lett. 103, 093902 (2009)}.
\bibitem{rute10} C.E. R\"uter, K. G. Makris, R. El-Ganainy, D. N. Christodoulides, M. Segev, and D. Kip, \href{https://doi.org/10.1038/nphys1515}{Nat. Phys. 6, 192 (2010)}.

\bibitem{pick13} J. Pickton and H. Susanto, \href{https://doi.org/10.1103/PhysRevA.88.063840} {Phys. Rev. A 88, 063840 (2013)}.



\bibitem{such11} S V Suchkov, B A Malomed, S V Dmitriev, Yu.\ S Kivshar, \href{https://doi.org/10.1103/PhysRevE.84.046609}{Phys. Rev. E 84, 046609 (2011)}.


\bibitem{kiri16} O.B.\ Kirikchi, A.A.\ Bachtiar and H.\ Susanto, \href{http://dx.doi.org/10.1155/2016/9514230}{Advances in Mathematical Physics 2016, 9514230 (2016)}.

\bibitem{alex12} N. V. Alexeeva, I. V. Barashenkov, A. A. Sukhorukov, and Y. S. Kivshar, Phys. Rev. A 85, 063837 (2012).

\bibitem{alex17} N.V. Alexeeva, I.V. Barashenkov, and Yu.S. Kivshar, New J. Phys. 19, 113032 (2017).



\bibitem{burl16} G.\ Burlak, S.\ Garcia-Paredes, and B.A.\ Malomed, Chaos 26, 113103 (2016). 

\bibitem{burl13} G. Burlak and B. A. Malomed, Phys. Rev. E 88, 062904 (2013). 



\bibitem{cout91} J.-L. Coutaz and M. Kull, \href{https://doi.org/10.1364/JOSAB.8.000095}{J. Opt. Soc. Am. B 8, 95 (1991)}.
\bibitem{smir06} E. Smirnov, C.E. R\"uter, M. Step\'ic, D. Kip, and V. Shandarov, \href{https://doi.org/10.1103/PhysRevE.74.065601}{Phys. Rev. E 74, 065601(R) (2006)}.


\bibitem{kusd16} R. Kusdiantara and H. Susanto, Phys. Rev. E 96, 062214 (2017).

\bibitem{cher16} A.\ Chernyavsky and D.E.\ Pelinovsky, J. Phys. A: Math. Theor. 49 (2016) 475201

\bibitem{dest17} E.Destyl, S.P.Nuiro, D.E.Pelinovsky and P.Poullet, Phys.\ Lett.\  A 381, 3884-3892 (2017)


\bibitem{li17} X.\ Li and Z.\ Yan, Chaos 27, 013105 (2017). 



\bibitem{kirr08} E.W. Kirr, P.G. Kevrekidis, E. Shlizerman, and M. I. Weinstein, SIAM J. Math. Anal., 40(2), 566-604 (2008).

\bibitem{kirr11} E. Kirr, P. G. Kevrekidis, and D. E. Pelinovsky, Commun. Math. Phys. 308, 795-844 (2011).

\bibitem{smer97} A. Smerzi, S. Fantoni, S. Giovanazzi, and S. R. Shenoy, Phys. Rev. Lett. 79, 4950 (1997).
\bibitem{ragh99} S. Raghavan, A. Smerzi, S. Fantoni, and S. R. Shenoy, Phys. Rev. A 59, 620633 (1999).

\bibitem{jian14} H. Jiang, H. Susanto, T. M. Benson, and K. A. Cliffe
Phys. Rev. A 89, 013828 (2014).

\bibitem{rodr13} A. S. Rodrigues, K. Li, V. Achilleos, P. G. Kevrekidis, D. J. Frantzeskakis, and C. M. Bender, \href{http://www.rrp.infim.ro/2013_65_1/art01Rodrigues.pdf}{Romanian Reports in Physics 65, 5-26 (2013)}.

\bibitem{hill06} M. Hiller, T. Kottos, and A. Ossipov, Phys. Rev. A, 73, 063625 (2006).

\bibitem{cart12} H. Cartarius and G. Wunner, Phys. Rev. A, 86, 013612 (2012);
H. Cartarius, D. Haag, D. Dast, and G. Wunner, J. Phys. A: Math. Theor., 45, 444008 (2012).


\bibitem{yang12} J. Yang, 
Stud. Appl. Math. 129, 133?162 (2012); 
Physica D 244, 50?67 (2013).


\bibitem{achi14} Achilleos V., Kevrekidis P.G., Frantzeskakis D.J., Carretero-Gonz\'alez R. (2014) 
In: Carretero-González R., Cuevas-Maraver J., Frantzeskakis D., Karachalios N., Kevrekidis P., Palmero-Acebedo F. (eds) Localized Excitations in Nonlinear Complex Systems. Nonlinear Systems and Complexity, vol 7. Springer, Cham.

\bibitem{li13} K.\ Li, P.G.\ Kevrekidis, D.J.\ Frantzeskakis, C.E.\ R\"uter and D.\ Kip, J. Phys. A: Math. Theor. 46, 375304 (2013)

\bibitem{carp03} A.\ Carpio and L.L.\ Bonilla, 
\href{https://doi.org/10.1137/S003613990239006X}{SIAM J.\ Appl.\ Math.\ 63, 1056-1082 (2003)}.

\bibitem{wolf12} M.\ Wolfrum, 
\href{http://dx.doi.org/10.1016/j.physd.2012.05.002}{Physica D 241, 1351-1357 (2012)}.


\bibitem{flac96} S. Flach and K. Kladko, \href{https://doi.org/10.1103/PhysRevE.54.2912}{Phys. Rev. E 54, 2912 (1996)}.

\bibitem{sagd88} R. S. Sagdeev, D. A. Usikov, and G. M. Zaslavski, \emph{Nonlinear
	Physics: from the Pendulum to Turbulence and Chaos} (Harwood Academic, Chur, Switzerland, 1988).

\bibitem{dawe13} J.H.P. Dawes and H. Susanto, \href{https://doi.org/10.1103/PhysRevE.87.032103}{Phys. Rev. E 87, 063202-7 (2013)}.

\bibitem{kozy13} G. Kozyreff and S. J. Chapman, 
Phys. Rev. Lett. 111, 054501 (2013).

\bibitem{king01} J. R. King and S. J. Chapman, 
European Journal of Applied Mathematics  12, 433-463 (2001).

	\end{thebibliography}
\end{document}